\documentclass[12pt]{iopart}

\usepackage{iopams}
\usepackage{graphicx}
\begin{document}

\title{Arbitrary quantum control of qubits in the presence of universal noise}

\author{Todd J. Green, Jarrah Sastrawan}

\address{ARC Centre for Engineered Quantum Systems, School of Physics\\ The University of Sydney, NSW 2006 Australia}

\author{Hermann Uys}
\address{National Laser Centre, Council for Scientific and Industrial Relations\\ Meiring Naude Road, Brummeria,Pretoria, South Africa}

\author{Michael J. Biercuk}
\address{ARC Centre for Engineered Quantum Systems, School of Physics\\ The University of Sydney, NSW 2006 Australia}
\ead{michael.biercuk@sydney.edu.au}

\begin{abstract}
We address the problem of deriving \emph{analytic} expressions for calculating universal decoherence-induced errors in qubits undergoing arbitrary, unitary, time-dependent quantum-control protocols. We show that the fidelity of a control operation may be expressed in terms of experimentally relevant spectral characteristics of the noise and of the control, over all Cartesian directions.  We formulate \emph{control matrices} in the time domain to capture the effects of piecewise-constant control, and convert them to generalized Fourier-domain filter functions.  These generalized filter functions may be derived for complex temporally modulated control protocols, accounting for susceptibility to rotations of the qubit state vector in three dimensions.  Taken together, we show that this framework provides a computationally efficient means to calculate the effects of universal noise on arbitrary quantum control protocols, producing results comparable to those obtained via time-consuming simulations of Bloch vector evolution. As a concrete example, we apply our method to treating the problem of dynamical decoupling incorporating realistic control pulses of arbitrary duration or form, including the replacement of simple $\pi$-pulses with complex dynamically corrected gates.
\end{abstract}

\maketitle

\section{Introduction}
A basic requirement for the realization of practical quantum coherent technologies, in particular quantum information processing (QIP) devices, is the capacity to efficiently manipulate quantum states with a high degree of precision. This prerequisite has given rise to the development of the field of Quantum Control Theory \cite{Tarn80, Tarn1983, Tarn2003, James2007, James2009, wiseman2010, d2007introduction}.  Significant research attention has been devoted to this discipline due to the remarkable promise of quantum coherent technologies for future applications including quantum computation, quantum-enabled sensing, and quantum-enhanced metrology.

One of the primary challenges being addressed by the research community is the understanding and mitigation of decoherence processes in quantum systems.  In these processes, uncontrolled interaction with the environment leads to randomization of a system's state, effectively destroying its ``quantumness.''  This phenomenon is especially important in quantum information settings where net error rates deep below fault-tolerance thresholds ($\sim10^{-4}$) are required in order to build scalable quantum computers \cite{Knill1998, Aliferis2008, NC}.


A variety of methods have been developed to characterize the performance of quantum control protocols, with an eye towards estimating error rates due to decoherence in realistic, experimentally relevant noise environments.  One of the most interesting, from a practical perspective, is the concept of \emph{spectral overlap} formalized by Kofman and Kurizki \cite{Kurizki2001, Kurizki2004}.  In this general approach, the net susceptibility of a given quantum control protocol to environmental noise is given by the overlap in frequency between the noise power spectral density and the spectral characteristics of the modulation imparted by the control. Such insights have been particularly important for the field of dynamical error suppression (DES), which seeks to provide error robustness to quantum hardware at the physical level.  These techniques address both implementation of quantum memory (dynamical decoupling) \cite{Ban1998, ViolaLloyd1998, ViolaKnill1999, Zanardi1999, Vitali1999} as well as nontrivial quantum logic gates \cite{MerrillBrown, ViolaKhod2009, ViolaKhod2009A} and exploit interference trajectories to effectively time-reverse the accumulation of error.

The concepts of the spectral overlap were expanded in the context of dynamical decoupling by Uhrig and Cywinski \cite{Uhrig2007, Cyw2008}, in attempting to understand the efficacy of these protocols in suppressing decoherence in Non-Markovian environments.  It was shown that, in considering DES broadly, the relevant control protocol could be thought of as a noise \emph{filter}, and the efficiency of a particular control sequence deduced simply by an examination of the relevant filter transfer function~\cite{mbad2011}. This general approach has been repeatedly validated in experiments using a variety of technologies - from trapped atomic ions to semiconductor spin qubits~\cite{Biercuk2009, Biercuk2009a, HansonDD, Yacoby2010, Medford2012}.  Remarkable agreement has been demonstrated between simple theoretical expressions for error derived from the so-called ``filter-function'' and experimentally measured noise power spectra - so much so that inverting this approach has allowed DES protocols to be used in noise spectroscopy~\cite{HansonMagnetometry, Bylander2011, Kotler2011}.

Despite these capabilities, there remains a significant gap in our understanding of how to efficiently calculate expected operational fidelities and error rates for complex quantum control protocols and for situations in which universal (i.e. multi-axis) time-dependent noise sources are present.  For instance, relatively little is known about how to account for the accumulation of error due to pulse non-idealities in dynamical decoupling sequences; the bang-bang limit is still widely assumed in analytic treatments.

Generally, while it is understood that spectral overlap techniques may be employed in order to evaluate error rates, it is not known how to efficiently produce analytic filter functions appropriate for such complex quantum control protocols.  This is largely due to that fact that the analytic complexity of deriving such functions grows significantly as soon as non-commuting operators appear in the control Hamiltonian - as would be the case in dynamical decoupling with nonzero-duration control pulses or other nontrivial control operations.

In this manuscript we address this challenge by developing a generalized method for evaluating the effects of universal, semiclassical decoherence on a quantum system undergoing an arbitrary unitary control protocol.  By adopting an approach in which the effect of the applied control is described by a three-dimensional control matrix \cite{Byrd2002, Pasini2008, Fauseweh2012}, the effect of universal decoherence on an effective spin-$1/2$ qubit is given a geometric interpretation.  In the presence of weak decohering noise, we show how all Cartesian contributions to the resulting operational fidelity may be calculated using noise power spectral densities along with the Fourier-space representations of the elements of the control matrix.    Further, we show how it is possible to simply construct generalizations of the filter function for arbitrary control protocols by assuming piecewise-constant controls.  As an example of the utility of this approach we study dynamical decoupling incorporating both realistic ``primitive'' $\pi$-pulses and more complicated dynamically corrected gates.  We are able to derive first order filter functions for the incorporation of arbitrary control pulses into a dynamical decoupling sequence and validate the performance of dynamically corrected gates in mitigating pulse errors in these sequences.

The remainder of the manuscript is organized as follows. In section~\ref{Sec:Theory} we first provide some relevant background theory before introducing a framework for quantum control in noisy environments and a geometric interpretation of decoherence, leading to the derivation of analytic expressions for the fidelity in section~\ref{Sec:Main}.  Section~\ref{Sec:Piecewise} then provides a concrete method to evaluate the control matrix (and hence the generalized filter functions) for the case of piecewise-constant control in the presence of universal and pure-dephasing noise.  To demonstrate the practical utility of our method this approach is applied to the problem of finite-pulse effects in dynamical decoupling in section~\ref{Sec:DD}.  Our framework permits us to efficiently separate out contributions to the filter function from the pulse \emph{locations} from effects arising from the pulse \emph{form} in a dephasing environment.  We follow with a discussion of the limits of the filter function approach and a comparison between this analytic framework and detailed numerics simulations in section~\ref{Sec:Limits}, before offering a conclusion.

\section{Quantum control in realistic environments}\label{Sec:Theory}

According to quantum theory, an initially coherent quantum system $S$ and its environment $E$ can, via physical interaction, become \emph{entangled}, forming a single composite quantum system $SE$. From the perspective of an observer, with the capacity to measure only $S$, coherence is effectively lost \cite{schlosshauer2008}. If unaddressed, this process of \emph{decoherence} can occur very rapidly, making the long term retention and accurate processing of information impossible.

Similar effects may be derived using a model in which environmental interaction is introduced via a fluctuating classical noise field. In this case, loss of coherence is understood as occurring via a process of randomization across an ensemble of identical quantum systems, each subject to a different realization of the noise. This semi-classical approach, often also referred to generically as decoherence, has the advantage of not requiring a detailed knowledge of the system-environment interaction Hamiltonian. Further, it has been shown repeatedly that such phenomenological noise models can accurately represent important aspects of experimental reality~\cite{Biercuk2009a}.

In this section, we outline a mathematical treatment of a controlled qubit affected by classical fields that serve to randomize the phase and probability amplitudes of the quantum state, while preserving the vector norm (i.e., we omit leakage from the computational state space).

\subsection{Single-qubit dynamics and the Bloch sphere}

Broadly speaking, a qubit is a quantum system that exhibits two possible outcomes for the measurement of a particular physical observable: the $z$-component of the intrinsic angular momentum of a spin-1/2 particle being the archetypal example. Letting $|0\rangle$ and $|1\rangle$ denote the states in which one or other of these outcomes are returned with certainty, and assuming maximal knowledge of the system, an arbitrary qubit state may be written as a linear combination
\begin{equation}\label{Eq:qubitstate}
|\psi\rangle=\cos(\vartheta/2)|0\rangle+e^{i\varphi}\sin(\vartheta/2)|1\rangle
\end{equation}
The angles $(\vartheta, \varphi)$ define a three-dimensional unit vector, the Bloch vector, that provides a convenient pictorial representation of the state \cite{NC}. Interpreting all possible states in this way, the two-dimensional qubit state space is mapped to a sphere of unit radius, called the Bloch sphere (figure \ref{Fig:Bloch1}).

The temporal evolution of the qubit is described by the propagator $U(t,0)$, a unitary operator that transforms some initial state $|\psi(0)\rangle$ to a final state $|\psi(t)\rangle=U(t,0)|\psi(0)\rangle$, for $t\geq0$. Within the Bloch sphere picture, this ``length-preserving'' transformation corresponds to a simple rotation of the qubit Bloch vector. The propagator satisfies the Schrodinger equation ($\hbar=1$)
\begin{equation}\label{Eq:se}
i\frac{d}{dt}U(t,0)=H(t)U(t,0)
\end{equation}
where $H(t)$ is the Hamiltonian operator, which we may assume to be traceless (i.e., $\Tr{\left(H(t)\right)}=0$),
for all $t$.  Disregarding irrelevant global phase factors, the set of all single qubit propagators, in conjunction with the usual operator product, forms the Lie group $SU(2)$ \cite{tung1985group}.

That a geometrical interpretation of the action of a propagator is possible in terms of a rotation of the Bloch vector is a consequence of the homomorphism (structure-preserving map) that may be constructed between $SU(2)$ and the group of three-dimensional rotation operators, $SO(3)$ \cite{tung1985group}. In short, the action of each propagator $U\in SU(2)$ (and its physically equivalent negative $-U$) on a single-qubit state $|\psi\rangle$ can be represented by the action of a rotation operator $R\in SO(3)$ on the Bloch vector corresponding to $|\psi\rangle$. To make this relationship explicit, we note that an arbitrary solution to \eref{Eq:se} may be written in the form $U(t,0)=\exp{[-i\theta(t)\hat{\bi{n}}(t)\boldsymbol{\sigma}/2]}$, where $\hat{\bi{n}}(t)=\left(n_{x}(t),n_{y}(t),n_{z}(t)\right)$ is a real row vector of unit length, $\boldsymbol{\sigma}=(\sigma_{x}, \sigma_{y}, \sigma_{z})^{T}$ is a column vector comprised of the Pauli spin operators, and $\theta(t)$ is a real function of time. The rotation operator in $SO(3)$ corresponding to $U(t,0)$ is that which rotates the Bloch vector through an angle $\theta(t)$, about an axis defined by $\hat{\bi{n}}(t)$. Thus, any single-qubit propagator may be thought of as acting to cause a simple rotation of the Bloch vector.

\subsection{Capturing the effects of noise}

To model the effect of noise on the capacity to control the evolution of a simple quantum system, we consider an ensemble of identically prepared noninteracting qubits evolving according to a Hamiltonian
\begin{equation}\label{Eq:ham0}
H(t)=H_{0}(t)+H_{c}(t)
\end{equation}
Here, the effect of the environment is modeled by the generalized noise Hamiltonian
\begin{equation}\label{Eq:ham}
H_{0}(t)=\boldsymbol{\beta}(t)\boldsymbol{\sigma}
\end{equation}
where $\boldsymbol{\beta}(t)=\left(\beta_{x}(t), \beta_{y}(t), \beta_{z}(t)\right)$ is a three-element row vector, each component of which is a random process modeling classical noise in one of the three spatial dimensions. Control over the state of the qubit can be achieved through the application of an external field, represented by a control Hamiltonian
\begin{equation}
H_{c}(t)=\bi{h}(t)\boldsymbol{\sigma}
\end{equation}
where the components of the vector $\bi{h}(t)=(h_{x}(t),h_{y}(t),h_{z}(t))$ describe the strength and direction of the control field as a function of time.

\begin{figure}[t]
      \centering
      \includegraphics[width=160mm]{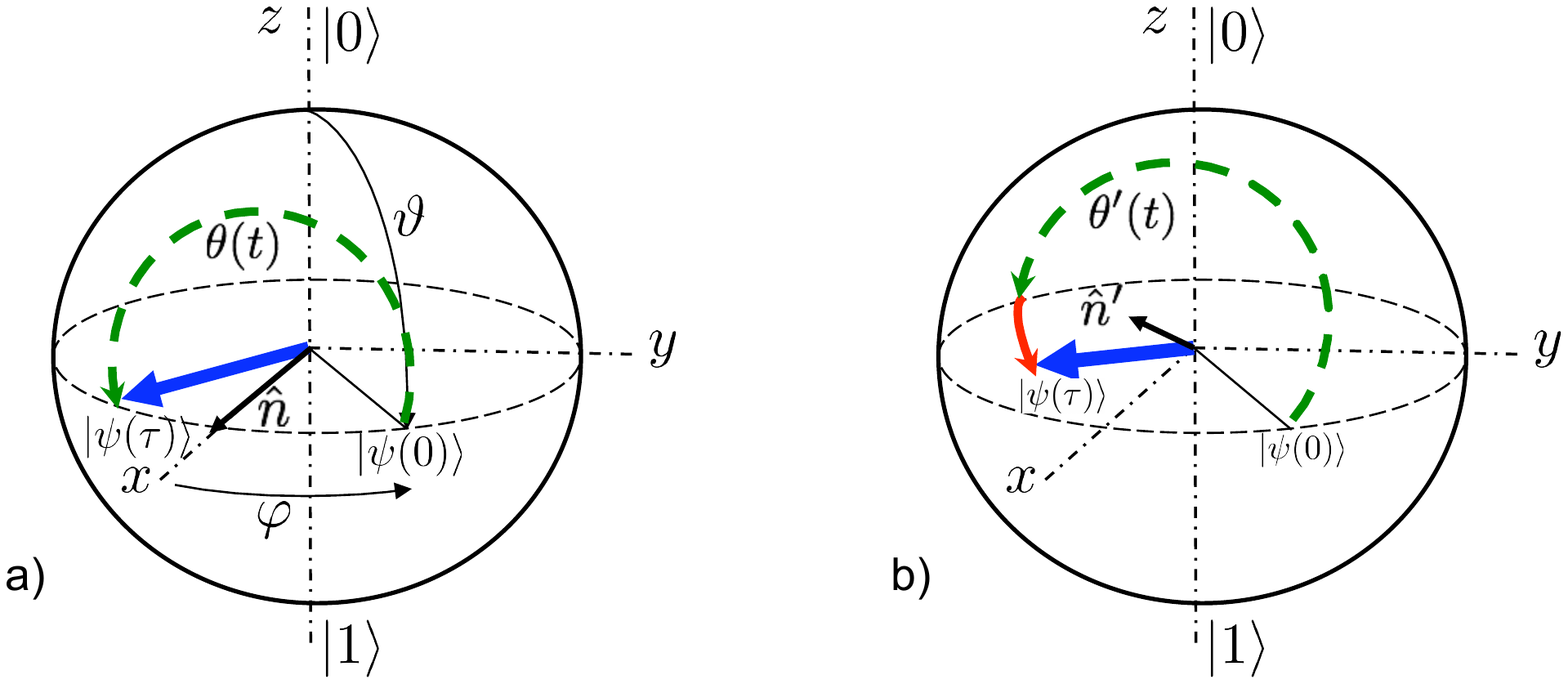}
      \caption{The Bloch sphere: showing a $\pi$-rotation about the x-axis, transforming an initial state $|\psi(0)\rangle$ to a final state $|\psi(t)\rangle$, for a) ideal and b) the noise-affected control operations.  The thick arrow indicates the final qubit state, the solid black arrow indicates the rotation axis, and the dotted path indicates the rotation angle.  In b) the rotation axis and angle are both changed by the presence of the noise.  Red extension of the path indicates overrotation in path $\theta'(t)$.  See text.}
      \label{Fig:Bloch1}
\end{figure}

The presence of the noise term in \eref{Eq:ham0} will, in general, affect adversely the ability to accurately `steer' the qubit state via the control Hamiltonian $H_{c}(t)$. As a simple illustration, consider the application of a constant control Hamiltonian $H_{c}(t)=\Omega\sigma_{x}/2$ to an otherwise isolated qubit, where $\Omega\equiv\pi/\tau$ for some positive real number $\tau$. Under these circumstances, the total Hamiltonian is independent of time and equation (\ref{Eq:se}) is easily integrated to give $U(\tau,0)=\exp{[-i\pi\sigma_{x}/2]}$. Thus, over a time interval $[0,\tau]$ the control field generates a rotation of the qubit Bloch vector through an angle of $\pi$ radians, about the $x$-axis (a `$\pi_{X}$-pulse'), as shown in figure \ref{Fig:Bloch1}a. If we allow for nonzero but constant `noise' by letting $\boldsymbol{\beta}(t)=(0,0,\beta_{z})$, for example, then $U(\tau,0)=\exp{[-i\pi(\sigma_{x}+\nu\sigma_{z})/2]}$, where $\nu\equiv2\beta_{z}/\Omega$. The propagator now describes a rotation through an angle $\theta'>\pi$, about an axis that is tilted away from the $xy$-plane (figure \ref{Fig:Bloch1}b). The noise therefore causes both a change in the angle of rotation and a shift of the rotational axis, so that the final qubit state will not be the one intended.

In general, for noise that is time-dependent, deriving a simple closed-form expression for the propagator $U(t)$ is not possible, as the Hamiltonian will generally not commute with itself at different times (from here on we omit the first argument of a propagator when it is zero). Therefore, rather than attempt to solve (\ref{Eq:se}) directly, we introduce the `control propagator' $U_{c}(t)$, defined as the solution to the \emph{noise-free} Schrodinger equation $idU_{c}(t)/dt=H_{c}(t)U_{c}(t)$. We may then write $U(t)=U_{c}(t)\tilde{U}(t)$, with the `error propagator' $\tilde{U}(t)$ capturing any deviation from the control evolution due to noise \cite{LidarKK1,KV2009a}. Substituting $U(t)=U_{c}(t)\tilde{U}(t)$ into (\ref{Eq:se}), and defining the `toggling frame' Hamiltonian
\begin{equation}\label{Eq:togham}
\tilde{H}_{0}(t)\equiv U^{\dag}_{c}(t)H_{0}(t)U_{c}(t)
\end{equation}
we find that $\tilde{U}(t)$ satisfies the modified Schrodinger equation
\begin{equation}\label{Eq:togse}
i\frac{d}{dt}\tilde{U}(t)=\tilde{H}_{0}(t)\tilde{U}(t)
\end{equation}
In the absence of noise, executing a target operation $Q$ over a time interval $[0,\tau]$ simply requires one to engineer a control Hamiltonian such that $U_{c}(\tau)=Q$.  More generally, we retain the requirement that $U_{c}(\tau)=Q$, so that $U(\tau)=Q\tilde{U}(\tau)$. The goal of dynamical error suppression is then to reduce $\tilde{U}(\tau)$ to the identity $I$.

\subsection{Geometric treatment of decoherence: The error vector and control matrix}

The error propagator $\tilde{U}(\tau)$, being an element of $SU(2)$, may be interpreted as a rotation of the qubit Bloch vector through some angle $2a(\tau)$, about an axis $\mathbf{\hat{a}}(\tau)$. Hence
\begin{equation}\label{Eq:errvec}
\tilde{U}(\tau)=\exp{\left[-i\mathbf{a}(\tau)\boldsymbol{\sigma}\right]}
\end{equation}
where $\mathbf{a}(\tau)\equiv a(\tau)\mathbf{\hat{a}}(\tau)$ is the real-valued `error vector' with components $a_{i}(\tau)$, for $i\in\{x,y,z\}$, and magnitude $|a(\tau)|=\left(\mathbf{a}(\tau)\mathbf{a}^{T}(\tau)\right)^{1/2}$.

From equations~\eref{Eq:togham}-\eref{Eq:errvec}, we see that the error vector is determined by the toggling frame Hamiltonian $\tilde{H}_{0}(t)$, which itself derives from the action of the control propagator $U_{c}(t)$ on $H_{0}(t)$. The aforementioned homomorphism between $SU(2)$ and $SO(3)$ enables us to identify $U_{c}(t)$ with a three-dimensional `control matrix' $\bi{R}(t)\in SO(3)$ \cite{Byrd2002}
\numparts
\begin{eqnarray}
\tilde{H}_{0}(t)&=\sum_{i=x,y,z}\beta_{i}(t)U^{\dag}_{c}(t)\sigma_{i}U_{c}(t)\\
&=\sum_{i,j=x,y,z}\beta_{i}(t)R_{ij}(t)\sigma_{j}\\
&=\boldsymbol{\beta}(t)\bi{R}(t)\boldsymbol{\sigma}\label{Eq:rotmat}
\end{eqnarray}
\endnumparts
Using the orthogonality of the Pauli operators with respect to the Hilbert-Schmidt inner product, the elements of $\bi{R}(t)$ are
\begin{equation}\label{Eq:5}
R_{ij}(t)\equiv[\bi{R}(t)]_{ij}=\Tr{\left(U_{c}^{\dag}(t)\sigma_{i}U_{c}(t)\sigma_{j}\right)}/2
\end{equation}
for $i,j\in\{x,y,z\}$.

The control matrix may be written compactly as $\bi{R}(t)\equiv(\bi{R}_{x}(t),\bi{R}_{y}(t),\bi{R}_{z}(t))^{T}$, where $\bi{R}_{i}(t)\equiv(R_{ix}(t),R_{iy}(t),R_{iz}(t))$ is a row vector comprised of the $i$-th row of the control matrix. Using this notation, (\ref{Eq:rotmat}) becomes
\begin{equation}\label{Eq:vecform}
\tilde{H}_{0}(t)=\sum_{i=x,y,z}\beta_{i}(t)\bi{R}_{i}(t)\boldsymbol{\sigma}
\end{equation}
We will use this form in below, where we seek to write the error vector in terms of the noise components and elements of the control matrix.

\subsection{Magnus expansion of the error vector}
An explicit expression for the error vector $\mathbf{a}(\tau)$ can be calculated by finding a solution to (\ref{Eq:togse}) that has the exponential form (\ref{Eq:errvec}). At the cost of expressing the exponent as an infinite series, the Magnus expansion gives the desired result \cite{Magnus, Blanes, Ng2011}. Specifically, given equation \eref{Eq:togse}, we may write $\tilde{U}(\tau)=\exp{[-i\Phi(\tau)]}$, where
\numparts
\begin{eqnarray}\label{Eq:mag1}
\Phi(\tau)=\sum_{\mu}^{\infty}\Phi_{\mu}(\tau)
\end{eqnarray}
The first three terms in the expansion are
\begin{eqnarray}\label{Eq:mag2}
\Phi_{1}(\tau)=\int_{0}^{\tau}dt\tilde{H}_{0}(t)\\
\Phi_{2}(\tau)=-\frac{i}{2}\int_{0}^{\tau}dt_{1}\int_{0}^{t_{1}}dt_{2}\left[\tilde{H}_{0}(t_{1}),\tilde{H}_{0}(t_{2})\right]\\
\Phi_{3}(\tau)=-\frac{1}{6}\int_{0}^{\tau}dt_{1}\int_{0}^{t_{1}}dt_{2}\int_{0}^{t_{2}}dt_{3}
\left\{\left[\tilde{H}_{0}(t_{1}),\left[\tilde{H}_{0}(t_{2}),\tilde{H}_{0}(t_{3})\right]\right]\right.\nonumber\\
\hspace{6.6cm}\left.+\left[\tilde{H}_{0}(t_{3}),\left[\tilde{H}_{0}(t_{2}),\tilde{H}_{0}(t_{1})\right]\right]\right\}
\end{eqnarray}
\endnumparts
with higher order terms involving increasingly complicated multiple integrals of nested commutators.

Substituting the toggling frame Hamiltonian \eref{Eq:vecform} into the Magnus expansion, and using the vector identity $[\mathbf{u}\boldsymbol{\sigma},\mathbf{v}\boldsymbol{\sigma}]=2i(\mathbf{u}\times\mathbf{v})
\boldsymbol{\sigma}$, for $\mathbf{u}$, $\mathbf{v}\in\mathbb{{R}}^3$, we find that we are able to write the error vector in the form of an infinite series
\numparts
\begin{eqnarray}\label{Eq:errser}
\mathbf{a}(\tau)=\sum_{\mu}^{\infty}\mathbf{a}_{\mu}(\tau)
\end{eqnarray}
where
\begin{eqnarray}\label{Eq:errvecmag}
\fl\hspace{2cm}
\mathbf{a}_{1}(\tau)=\sum_{i=x,y,z}\int^{\tau}_{0}dt\beta_{i}(t)\mathbf{R}_{i}(t)\label{Eq:errvecmaga}\\
\fl\hspace{2cm}
\mathbf{a}_{2}(\tau)=\sum_{i,j=x,y,z}\int^{\tau}_{0}dt_{1}\int^{t_{1}}_{0}dt_{2}
\beta_{i}(t_{1})\beta_{j}(t_{2})\tilde{\bi{R}}_{ij}(t_{1},t_{2})\label{Eq:errvecmagb}\\
\fl\hspace{2cm}
\mathbf{a}_{3}(\tau)=\frac{2}{3}\sum_{i,j,k=x,y,z}\int_{0}^{\tau}dt_{1}\int_{0}^{t_{1}}dt_{2}\int_{0}^{t_{2}}dt_{3}
\beta_{i}(t_{1})\beta_{j}(t_{2})\beta_{k}(t_{3})\tilde{\bi{R}}_{ijk}(t_{1},t_{2},t_{3})\label{Eq:magerr2}
\end{eqnarray}
\endnumparts
Here, we've introduced the vectors
$
\tilde{\bi{R}}_{ij}(t_{1},t_{2}))\equiv\left[\bi{R}_{i}(t_{1})\times\bi{R}_{j}(t_{2})\right]
$
and
$
\tilde{\bi{R}}_{ijk}(t_{1},t_{2},t_{3})\equiv\left[\bi{R}_{i}(t_{1})\times
\left[\bi{R}_{j}(t_{2})\times\bi{R}_{k}(t_{3})\right]\right]
+\left[\bi{R}_{k}(t_{3})\times\left[\bi{R}_{j}(t_{2})\times\bi{R}_{i}(t_{1})\right]\right]
$.
Generally, the $n$-th order term is an $n$-fold integral over the sum of all possible products of the form $\beta_{i_{1}}(t_{1})\beta_{i_{2}}(t_{2})...\beta_{i_{n}}(t_{n})
\tilde{\bi{R}}_{i_{1}i_{2}...i_{n}}(t_{1},t_{2},...,t_{n})$, where the vector $\tilde{\bi{R}}_{i_{1}i_{2}...i_{n}}(t_{1},t_{2},...,t_{n})$ is a sum of multiple vector cross products of rows of the control matrix $\bi{R}(t)$, evaluated at times $t_{1},t_{2},...,t_{n}$.

Given \eref{Eq:errser}-\eref{Eq:magerr2}, we have a series expansion for the error vector in which the components of the noise vector and the elements of the control matrix appear explicitly. By defining an appropriate metric for the fidelity of a control operation, in terms of the error vector, we can use these results to evaluate the impact of noise on an arbitrary control operation on an ensemble of identical non-interacting qubits.

\section{Operational fidelity}\label{Sec:Main}

One way of measuring how well a given propagator $V_{1}$ approximates a target operation $V_{2}$ is to take the Hilbert-Schmidt inner product
$
(V_{2},V_{1})\equiv\frac{1}{2}\Tr{(V_{2}^{\dag}V_{1})}
$
which, by analogy with the usual inner product of state vectors, effectively measures the `overlap' between the two operators. Taking the square modulus of the inner product of $U(\tau)=Q\tilde{U}(\tau)$ and $Q$ gives $\mathcal{F}(\tau)\equiv\frac{1}{4}|\Tr(\tilde{U}(\tau))|^{2}$. This metric serves to quantify the accuracy of a control operation subject to a particular realization of the noise vector $\boldsymbol{\beta}(t)$. Experimentally, however, it is usually only the ensemble average (denoted by the angular brackets $\langle...\rangle)$ of the $\mathcal{F}(\tau)$, taken over all realizations of the noise Hamiltonian (\ref{Eq:ham}), that is measured. We therefore refer to
\begin{equation}\label{Eq:avfid}
\mathcal{F}_{av}(\tau)\equiv\frac{1}{4}\langle|\Tr(\tilde{U}(\tau))|^{2}\rangle
\end{equation}
as determining the `fidelity' of a particular control operation.

With the error propagator written in terms of the error vector \eref{Eq:errvec}, and suppressing the explicit $\tau$-dependence, one finds
\begin{equation}\label{Eq:fid}
\mathcal{F}_{av}=\frac{1}{2}\left[\left\langle\cos{(2a)}\right\rangle+1\right]
\end{equation}
so that, for each qubit in the ensemble, the fidelity is determined by the magnitude of the angle through which the noise causes the control operation to be \emph{rotated} away from the target operation $Q$.  Expanding the cosine term in a Taylor series we obtain
\begin{equation}
\mathcal{F}_{av}=\frac{1}{2}\left[1+\sum_{m=0}^{\infty}(-1)^{m}\frac{2^{2m}}{(2m)!}\langle a^{2m}\rangle\right]
\end{equation}
Considering the first nontrivial term ($m=1$) in the expansion, recalling that $a^2=\mathbf{a}\mathbf{a}^{T}$, and using the Magnus expansion of the error vector \eref{Eq:errvecmag}, we may write
\begin{eqnarray}
\fl
\hspace{1cm}\langle a^2\rangle=\sum_{\mu\nu}\langle\mathbf{a}_{\mu}\mathbf{a}_{\nu}^{T}\rangle=\left[\langle a_{1}^{2}\rangle+\langle a_{2}^{2}\rangle+...+2\left(\langle\mathbf{a}_{1}\mathbf{a}_{2}^{T}\rangle
+\langle\mathbf{a}_{1}\mathbf{a}_{3}^{T}\rangle+\langle\mathbf{a}_{2}\mathbf{a}_{3}^{T}\rangle+...\right)\right]
\end{eqnarray}
where $a_{1}^{2}\equiv\mathbf{a}_{1}\mathbf{a}_{1}^{T}$, and $\mu, \nu$ are indices indicating the order of the Magnus expansion. Calculating the terms $\langle a^{2m}\rangle$ for $m>1$ in a similar way, we arrive at a series expansion for the fidelity
\begin{equation}\label{Eq:fidexp}
\mathcal{F}_{av}=1-\langle a_{1}^{2}\rangle-2\langle \mathbf{a}_{1}\mathbf{a}^{T}_{2}\rangle-\left[\langle a_{2}^{2}\rangle+2\langle \mathbf{a}_{1}\mathbf{a}^{T}_{3}\rangle-\frac{\langle a_{1}^{4}\rangle}{3}\right]+...
\end{equation}
with $a_{2}^{2}\equiv\mathbf{a}_{2}\mathbf{a}_{2}^{T}$ and $a_{1}^{4}\equiv(\mathbf{a}_{1}\mathbf{a}_{1}^{T})^{2}$.

 The fidelity is thus expressed explicitly in terms of noise correlations and the control matrix. For instance, the second term in \eref{Eq:fidexp}
\numparts
\begin{eqnarray}\label{Eq:firstord}
\langle a_{1}^{2}\rangle&=\sum_{i,j=x,y,z}\int^{\tau}_{0}dt_{2}\int^{\tau}_{0}dt_{1}
\langle\beta_{i}(t_{1})\beta_{j}(t_{2})\rangle \bi{R}_{i}(t_{1})\bi{R}^{T}_{j}(t_{2})\\
&=\sum_{i,j,k=x,y,z}\int^{\tau}_{0}dt_{2}\int^{\tau}_{0}dt_{1}
\langle\beta_{i}(t_{1})\beta_{j}(t_{2})\rangle R_{ik}(t_{1})R_{jk}(t_{2})
\end{eqnarray}
\endnumparts
contains all two-point noise cross-correlation functions $\langle\beta_{i}(t_{1})\beta_{j}(t_{2})\rangle$, for $i,j\in\{x,y,z\}$, while the third contains all those evaluated at three time points, and the terms in square brackets all involve four-point correlation functions (this is determined by the sum of subscript indices, as they indicate the expansion-order of the error vector in~\eref{Eq:errvecmag}-\eref{Eq:magerr2}.  For noise that is weak, the general trend will be for terms involving higher-order correlation functions to have less of an effect on the fidelity.  Explicit forms for these terms, in the case that the components of the noise vector are independent Gaussian processes, may be found in the Appendix of this paper and, for purely dephasing noise (i.e. $H_{0}(t)=\beta_{z}(t)\sigma_{z}$), in~\cite{Green2012}.

\subsection{Moving to the Fourier domain}
The individual terms in the series expansion of the fidelity \eref{Eq:fidexp} rely on time-domain correlation and cross-correlation functions and convolution with the multidimensional control matrix. Using the properties of the Fourier transform, it is experimentally more convenient to rewrite these terms in the frequency domain. Generally, we can define the Fourier transform $\mathcal{S}_{i_{1}...i_{n}}(\omega_{1},...,\omega_{n})$ of an $n$-point cross-correlation function via
\begin{equation}
\fl\hspace{.5cm}
\langle\beta_{i_{1}}(t_{1})\beta_{i_{2}}(t_{2})...\beta_{i_{n}}(t_{n})\rangle
\equiv\frac{1}{(2\pi)^{n}}\int d\omega_{1}...\int d\omega_{n} \mathcal{S}_{i_{1}...i_{n}}(\omega_{1},...,\omega_{n})e^{i(\omega_{1}t_{1}+...+\omega_{n}t_{n})}
\end{equation}
The fidelity \eref{Eq:fidexp} can then be rewritten as
\begin{equation}\label{Eq:fidexpfreq}
\fl\hspace{.5cm}
\mathcal{F}_{av}=1-\sum_{n=2}^{\infty}
\left\{\frac{1}{(2\pi)^{n}}\sum_{i_{1}...i_{n}}\int d\omega_{1}...\int d\omega_{n}
\mathcal{S}_{i_{1}...i_{n}}(\omega_{1},...,\omega_{n})\mathcal{R}{i_{1}...i_{n}}(\omega_{1},...,\omega_{n})
\right\}
\end{equation}
where $\mathcal{R}_{i_{1}...i_{n}}(\omega_{1},...,\omega_{n})$ is determined solely by the control matrix.

For noise that is sufficiently weak, the $n=2$ term in \eref{Eq:fidexpfreq} will be the dominant contributor to fidelity loss in the short term. If the noise is also \emph{wide sense stationary} then the two-point cross-correlation functions depend only on the time difference $t_{2}-t_{1}$, so that
\begin{equation}
\langle\beta_{i}(t_{1})\beta_{j}(t_{2})\rangle=\frac{1}{2\pi}\int_{-\infty}^{\infty}d\omega
S_{ij}(\omega)e^{i\omega(t_{2}-t_{1})}
\end{equation}
where $S_{ij}(\omega)$ is the cross-power spectral density between the random variables
$\beta_{i}(t)$ and $\beta_{j}(t)$. We then have
\begin{equation}\label{Eq:result2}
\mathcal{F}_{av}\simeq1-\frac{1}{2\pi}\sum_{i,j,k=x,y,z}\int_{-\infty}^{\infty}\frac{d\omega}
{\omega^2}S_{ij}(\omega)R_{jk}(\omega)R^{*}_{ik}(\omega)
\end{equation}
where
\begin{equation}\label{Eq:result3}
R_{ij}(\omega)\equiv-i\omega\int_{0}^{\tau}dtR_{ij}(t)e^{i\omega t}
\end{equation}
are the elements of the control matrix in the frequency domain (again we suppress the explicit dependence on the total time $\tau$).  The form of~(\ref{Eq:result2}) is reminiscent of the filter function introduced in previous literature on dynamical error suppression, to which we will return later. Thus, in the weak noise regime (to be defined in Section ??, for Gaussian noise), it is possible to understand the fidelity of a qubit's unitary evolution in terms of spectral properties of the noise and the control~\cite{Kurizki2001,Uhrig2007}.

In the next section we will show how the control matrix may be evaluated for the paradigmatic case of a piecewise constant control sequence. In doing so, we explicitly demonstrate the connection between our generalized approach and the well known filter-function formalism.

\section{Evaluating the control matrix}\label{Sec:Piecewise}
In general, for arbitrary continuous-time modulation of $H_{c}(t)$, the control matrix can be evaluated only approximately.  However, under the simplifying assumption of piecewise-constant control it becomes possible to find an exact analytic form. The result can be applied to dynamical decoupling (state preservation) sequences, as well as a variety of nontrivial control operations such as piecewise-defined dynamically corrected gates and composite pulses \cite{levitt, MerrillBrown}.

\subsection{Universal noise}
\begin{figure}[t]
      \centering
      \includegraphics[width=160mm]{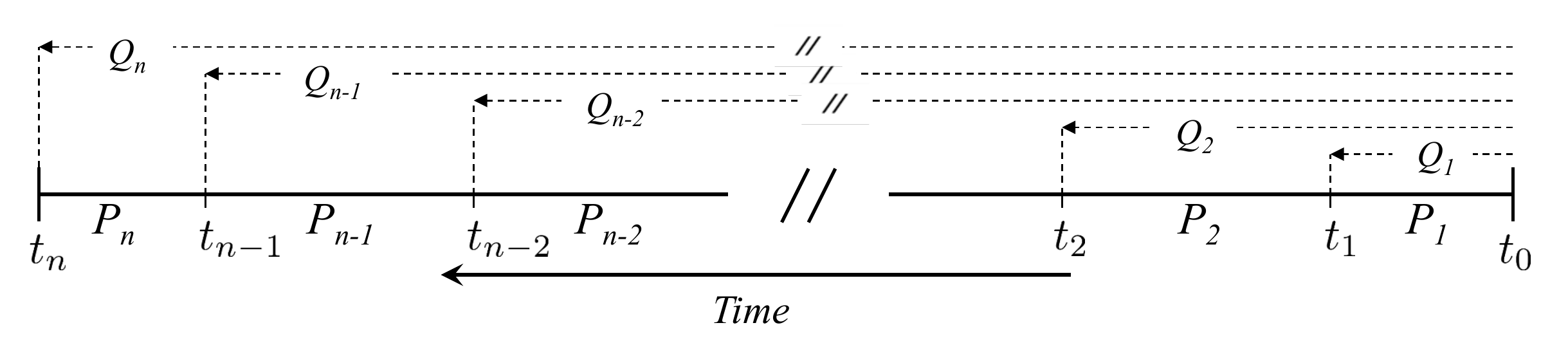}
      \caption{Explanatory diagram for a piecewise-defined control sequence. During each interval $[t_{l-1},t_{l}]$, a control operation $P_{l}$ is executed. The operators $Q_{l}$ are the control propagator evaluated at the start of the $l$-th pulse, for $l=1,2,..,n$.}
      \label{Fig:Piecewise}
\end{figure}

For a sequence of $n$ consecutive unitary control operations (or control `pulses') $P_{1},...,P_{n}$, we divide the total sequence time $\tau$ into $n$ corresponding time intervals, such that the $l$-th operation $P_{l}$ is executed over the interval $[t_{l-1},t_{l}]$, i.e., $P_{l}=U_{c}(t_{l},t_{l-1})$, for $l\in\{1,2,...,n\}$, where $t_{n}\equiv\tau$ and $t_{0}\equiv0$ (figure \ref{Fig:Piecewise}). Letting $P_{0}\equiv I$, and defining the cumulative operators $Q_{l}\equiv P_{l}P_{l-1}...P_{0}$ so that $Q\equiv Q_{n}$, the elements of the control matrix \eref{Eq:5} take the form
\begin{equation}\label{Eq:seq}
R_{ij}(t)=\frac{1}{2}\sum_{l=1}^{n}G^{(l)}(t)
\Tr{\left(Q_{l-1}^{\dag}U_{c}^{\dag}(t,t_{l-1})\sigma_{i}
U_{c}(t,t_{l-1})Q_{l-1}\sigma_{j}\right)}
\end{equation}
where the function $G^{(l)}(t)\equiv\Theta[t-t_{l-1}]-\Theta[t-t_{l}]$ has unit value within the $l$-th time interval and is zero elsewhere.

Equation \eref{Eq:seq} can be rewritten in terms of the individual control matrices $\bi{R}^{P_{l}}(t-t_{l-1})$ for each of the $n$ operations $P_{l}$ in the sequence, defined relative to the pulse start times $t_{l-1}$, by letting
\begin{equation}\label{Eq:pulsematrix}
U_{c}^{\dag}(t,t_{l-1})\sigma_{i}U_{c}(t,t_{l-1})=\sum_{j=x,y,z}R_{ij}^{P_{l}}(t-t_{l-1})\sigma_{j}
\end{equation}

\noindent for $i\in \{x,y,z\}$. Again, the components of $\bi{R}^{P_{l}}(t-t_{l-1})$ are obtained from the Hilbert-Schmidt inner product
\begin{equation}\label{Eq:pulsecomp}
R_{ij}^{P_{l}}(t-t_{l-1})\equiv[\bi{R}^{P_{l}}(t-t_{l-1})]_{ij}=\frac{1}{2}\Tr{\left(U_{c}^{\dag}(t,t_{l-1})\sigma_{i}U_{c}(t,t_{l-1})\sigma_{j}\right)}
\end{equation}for $i,j\in\{x,y,z\}$. Substituting \eref{Eq:pulsematrix} into \eref{Eq:seq}, and using the linearity of the trace operation, we find that we can write the control matrix for the pulse sequence in the compact form
\begin{equation}\label{Eq:seqcontmat}
\bi{R}(t)=\sum_{l=1}^{n}G^{(l)}(t)\bi{R}^{P_{l}}(t-t_{l-1})\boldsymbol{\Lambda}^{(l-1)}
\end{equation}
where the matrix $\boldsymbol{\Lambda}^{(l-1)}$ has components
\begin{equation}\label{Eq:Rswitchmatrix}
\Lambda_{ij}^{(l-1)}\equiv[\boldsymbol{\Lambda}^{(l-1)}]_{ij}=\frac{1}{2}\Tr{\left(Q_{l-1}^{\dag}\sigma_{i}Q_{l-1}\sigma_{j}\right)}
\end{equation}
for $i,j\in\{x,y,z\}$. Using \eref{Eq:result3} to define the control matrix in the frequency domain, we have
\numparts
\begin{equation}\label{Eq:Rseq2w}
\bi{R}(\omega)=\sum_{l=1}^{n}e^{i\omega t_{l-1}}\bi{R}^{P_{l}}(\omega)\boldsymbol{\Lambda}^{(l-1)}
\end{equation}
where
\begin{equation}\label{Eq:Rrw}
\bi{R}^{P_{l}}(\omega)\equiv-i\omega\int_{0}^{t_{l}-t_{l-1}}dte^{i\omega t}\bi{R}^{P_{l}}(t)
\end{equation}
\endnumparts
is the frequency domain control matrix for the $l$-th pulse, $P_{l}$.

With the above expressions and a knowledge of the statistical properties of the noise we can, in combination with the results of section~\ref{Sec:Main}, find an approximate value for the fidelity of an arbitrary piecewise-defined control sequence in a weak noise environment. This approach is straightforward to implement by hand or via numerics.  Ultimately the computational efficiency of calculating the control matrix and spectral overlap integrals far exceeds that of brute-force numerical simulations of the evolution of the Bloch vector~\cite{Green2012}.

\subsection{Dephasing noise} \label{Sec:Dephasing}

In many of the physical systems with the potential for use in nascent quantum information technologies, the characteristic single-qubit dephasing time $T_{2}$ is considerably smaller than the relaxation time $T_{1}$, and governed by different physical processes. It is important, therefore, that we understand the effects of purely dephasing noise on single-qubit control. To this end, we consider a model system in which the noise vector $\boldsymbol{\beta}(t)$ has only a $z$-component and the noise Hamiltonian (\ref{Eq:ham}) reduces to
\begin{equation}\label{Eq:depham}
H_{0}(t)=\beta_{z}(t)\sigma_{z}
\end{equation}
In this context, only the third row of the control matrix is required to calculate the fidelity, and we refer to the corresponding row vector $\bi{R}_{z}(t)\equiv(R_{zx}(t),R_{zy}(t),R_{zz}(t))$  as the `control vector'. From \eref{Eq:result2}, the lowest-order contribution of noise to the fidelity \eref{Eq:fidexp} is captured by the term
\begin{equation}
\langle a_{1}^{2}\rangle=\frac{1}{2\pi}\int_{-\infty}^{\infty}\frac{d\omega}
{\omega^2}S_{z}(\omega)F^{(1)}_{z}(\omega)
\end{equation}
where the first order dephasing `filter function' $F^{(1)}_{z}(\omega)$ is simply the square modulus of the frequency domain control vector
\begin{equation}\label{Eq:FF}
F^{(1)}_{z}(\omega)\equiv
\sum_{i=x,y,z}|R_{zi}(\omega)|^{2}
\end{equation}
If we further assume that the noise is Gaussian, with a mean value of zero, then all correlation functions $\langle\beta_{z}(t_1)...\beta_{z}(t_{n})\rangle$ for which $n$ is odd are equal to zero. The Gaussian moment theorem also allows us to write all remaining higher-order correlations in terms of only the simplest two-point correlation function (see Appendix). In the frequency domain, the statistical properties of the noise are then captured entirely by the power spectral density $S_{z}(\omega)$, and the fidelity \eref{Eq:fidexp} can be written as the infinite sum \cite{Green2012}
\begin{eqnarray}\label{Eq:fidspec}
\fl
\hspace{1cm}
\mathcal{F}_{av}=1-\left[\frac{1}{2\pi}\int_{-\infty}^{\infty}\frac{d\omega}{\omega^2}S_{z}(\omega)F^{(1)}_{z}(\omega)\nonumber\right.\\
\hspace{1.6cm}\left.+\frac{1}{(2\pi)^{2}}\int_{-\infty}^{\infty}\frac{d\omega}{\omega^2}S_{z}(\omega)\int_{-\infty}^{\infty}\frac{d\omega'}{\omega'^2}
S_{z}(\omega')F^{(2)}_{zz}(\omega,\omega')+...\right]
\end{eqnarray}
Here, the effect of the control on the qubit fidelity is described by the functions $F^{(p)}_{z...}$, for $p=1,2,...$, which depend only on the control vector $\bi{R}_{z}$. The complexity of the filter functions increases rapidly with increasing order.  However, $F^{(1)}_{z}(\omega)$ can  be quite simple in form and reasonably straightforward to calculate. In the weak noise regime we ignore the higher order terms in (\ref{Eq:fidspec}) and write
\numparts
\begin{eqnarray}\label{Eq:approx}
\mathcal{F}_{av}(\tau)\simeq\frac{1}{2}\left\{1+\exp{\left[-\chi(\tau)\right]}\right\}
\end{eqnarray}
where the rate of fidelity decay is given by
\begin{eqnarray}\label{Eq:approx2}
\chi(\tau)\equiv\langle a_{1}^{2}\rangle=\frac{1}{\pi}\int_{0}^{\infty}\frac{d\omega}
{\omega^2}S_{z}(\omega)F^{(1)}_{z}(\omega)
\end{eqnarray}
\endnumparts
Thus, the decay rate is determined by a simple overlap integral between the power spectrum of the noise and a ``filter function'' representing the control in the Fourier domain. This result is similar to the familiar expression for the coherence of a dynamical decoupling sequence in the case of unbounded controls~\cite{Uhrig2007} (see also section \ref{Sec:DD}). The difference here being that the filter function $F^{(1)}_{z}$ captures only the lowest order nontrivial effect of the control on the fidelity in a dephasing environment, rather than providing an exact solution in the bang-bang limit.

For a piecewise-defined control sequence, subject to purely dephasing noise, equation~\eref{Eq:seq} reduces to the following expression for the components of the control vector
\begin{equation}\label{Eq:seqdeph}
R_{zi}(t)=\frac{1}{2}\sum_{l=1}^{n}G^{(l)}(t)
\Tr{\left(Q_{l-1}^{\dag}U_{c}^{\dag}(t,t_{l-1})\sigma_{z}
U_{c}(t,t_{l-1})Q_{l-1}\sigma_{i}\right)}
\end{equation}
Similarly, \eref{Eq:Rseq2w} and \eref{Eq:Rrw} become
\numparts
\begin{equation}\label{Eq:seq2w}
\bi{R}_{z}(\omega)=\sum_{l=1}^{n}e^{i\omega t_{l-1}}\bi{R}_{z}^{P_{l}}(\omega)\boldsymbol{\Lambda}^{(l-1)}
\end{equation}
and
\begin{equation}\label{Eq:rw}
\bi{R}_{z}^{P_{l}}(\omega)\equiv-i\omega\int_{0}^{t_{l}-t_{l-1}}dte^{i\omega t}\bi{R}_{z}^{P_{l}}(t)
\end{equation}
\endnumparts
respectively, where the components of the $l$-th pulse control vector are
\begin{equation}\label{Eq:pulsert}
R_{zj}^{P_{l}}(t-t_{l-1})=\frac{1}{2}\Tr{\left(U_{c}^{\dag}(t,t_{l-1})\sigma_{z}U_{c}(t,t_{l-1})\sigma_{j}\right)}
\end{equation}
for $j\in\{x,y,z\}$.

If we assume that during each of the $n$ time intervals comprising the sequence the applied control effects a steady rotation of the qubit Bloch vector through an angle $\theta_{l}$, about an axis $\hat{\bi{n}}_{l}=\cos{(\phi_{l})}\hat{x}+\sin{(\phi_{l})}\hat{y}$ in the $xy$-plane, then for $t\in[t_{l-1},t_{l}]$
\begin{equation}
U_{c}(t,t_{l-1})=\exp{\left[\frac{-i}{2}\Omega_{l}\left(t-t_{l-1}\right)\sigma_{\phi_{l}}\right]}
\end{equation}
where $\Omega_{l}\equiv\theta_{l}/(t_{l}-t_{l-1})$ is the rotation rate and $\sigma_{\phi_{l}}\equiv\cos{(\phi_{l})}\sigma_{x}+\sin{(\phi_{l})}\sigma_{y}$. From \eref{Eq:pulsert} and \eref{Eq:rw} we can show that the frequency domain control vector for the $l$-th operation has components
\begin{equation}\label{Eq:pulsew}
\fl
R_{zj}^{P_{l}}(\omega)=\frac{\omega}{\omega^2-\Omega_{l}^{2}}\left\{\delta_{zj}\left[i\Omega_{l}g_{l}(\omega)-\omega f_{l}(\omega)\right]
+\left[\Omega_{l}f_{l}(\omega)-i\omega g_{l}(\omega)\right]\Tr{(\sigma_{\phi_{l}}\sigma_{z}\sigma_{j})}\right\}
\end{equation}
where $f_{l}(\omega)\equiv e^{i\omega(t_{l}-t_{l-1})}\cos{(\theta_{l})}-1$ and $g_{l}(\omega)\equiv e^{i\omega(t_{l}-t_{l-1})}\sin{(\theta_{l})}$. Although \eref{Eq:pulsew} has been derived assuming rotations take place about axes in the $xy$-plane only, periods of free evolution (no rotation) and $z$-axis rotations are easily accounted for by setting $\theta_{l}=0$ and $\Omega_{l}=0$ in \eref{Eq:pulsew}, whenever one of these operations occurs in the sequence. This is possible simply because, for dephasing noise, both of these operations commute with the noise operator.

With these analytical tools, the process for finding the operational fidelity of an arbitrary piecewise-constant control sequence in the presence of dephasing noise may be summarized as follows:

\begin{enumerate}
\item Calculate the matrix defined in equation~\eref{Eq:Rswitchmatrix} for each step in the chain of control operations;
\item Substitute the above result and equation~\eref{Eq:pulsew} into~\eref{Eq:seq2w};
\item Repeat for all time steps $l\in\{1,...,n\}$ to find the control vector in the Fourier domain;
\item Find the first-order filter function from~\eref{Eq:FF};
\item Use~\eref{Eq:approx} and~\eref{Eq:approx2} to calculate approximate expression for the fidelity valid for weak dephasing noise.
\end{enumerate}

In the next section we employ this approach to study the impact of pulse imperfections on the performance of dynamical decoupling sequences used in quantum memory.

\section{Pulse effects in dynamical decoupling sequences}\label{Sec:DD}

\begin{figure}[t]
      \centering
      \includegraphics[width=160mm]{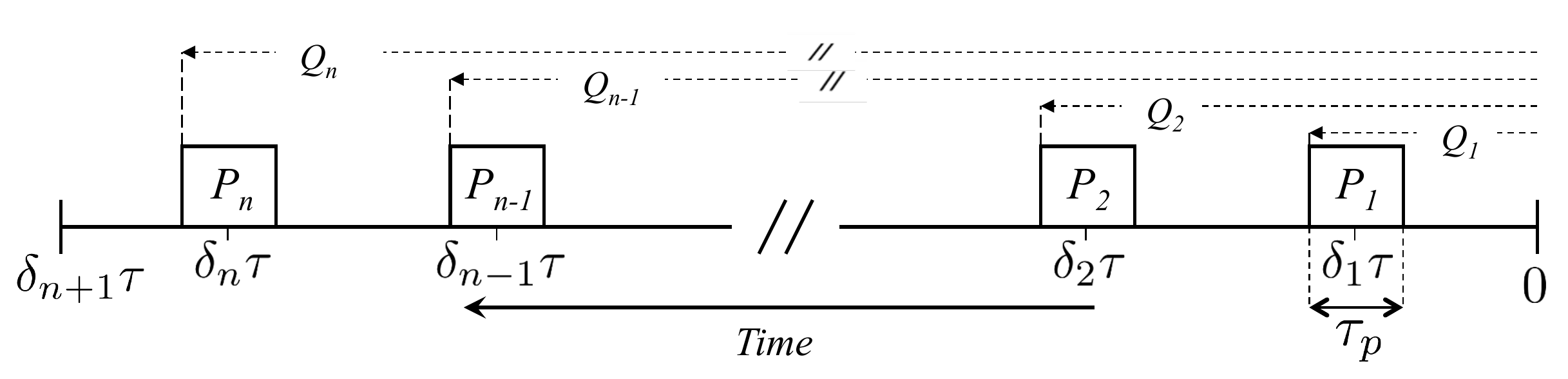}
      \caption{Dynamical decoupling sequence for $n$ identical pulses, each of width $\tau_{P}$, centered at times $\delta_{l}\tau,$ for $l=1,2,..,n$ }
      \label{Fig:DDsequence}
\end{figure}

Dynamical decoupling (DD) sequences are control protocols aimed at implementing the identity operation ($Q=I$), and thus preserving the initial state of the qubit. In the limit of instantaneous control pulses it has been shown that, for both semi-classical dephasing noise and quantum spin-boson models, it is possible to write the qubit coherence at the end of an arbitrary DD sequence, given preparation of the system in a superposition of eigenstates of $\sigma_{z}$, as $W=e^{-\chi(\tau)}$ \cite{Uhrig2008, Biercuk2009}.  In this expression, the effects of both the noisy dephasing environment and the control modulation are contained in the overlap integral
\begin{equation}\label{Eq:chi}
\chi(\tau)=\frac{1}{\pi}\int\limits _{0}^{\infty}
\frac{d\omega}{\omega^{2}}S_{z}(\omega)F_{z}(\omega)d\omega.
\end{equation}
The statistical properties of the noise in the Fourier domain are captured via the noise power spectrum $S_{z}(\omega)$, and the action of the control sequence via the filter function $F_{z}(\omega)$. We note that (\ref{Eq:chi}) is formally identical to (\ref{Eq:approx2}) above, with the exception that, because the control pulses are assumed to be unbounded in strength with vanishing duration (the so-called ``bang-bang'' approximation), the filter function in (\ref{Eq:chi}) is exact. The bang-bang approximation is, however, unphysical and has been shown in experiment to neglect important contributions to the net error arising from realistic pulses of nonzero width (see figure~\ref{Fig:DDsequence}).

In this section we apply the techniques we have developed to evaluate the fidelity of dynamical decoupling sequences when considering physical constraints - such as the incorporation of realistic control pulses - in a purely dephasing noise environment.

\subsection{Separating pulses from the pulse sequence}
Beyond the bang bang limit, the effectiveness of a DD sequence will depend on both the temporal spacing and form of the pulses used. To find optimal error-suppressing sequences for a given environment or physical system, either of these variables may be changed independently of the other, and these differences should be easily separated in an analytical framework. We are therefore motivated to express the sequence control vector $\bi{R}_{z}(\omega)$ in terms that can be separated into those dependent only on pulse \emph{location} and those dependent on pulse \emph{type} (expressed in terms of the individual \textit{pulse} control vectors $\bi{R}_{z}^{P_{l}}(\omega)$).

As in section \ref{Sec:Piecewise}, we start with a piecewise constant sequence of $n$ pulses, however, to model a DD sequence we allow for a period of free evolution (i.e., no control operation) both before and after each pulse, so that immediately following the $l$-th pulse, we have $Q_{l}=P_{l}IP_{l-1}I...IP_{0}$, for $l\in\{1,2,...,n\}$. Within the total sequence time $\tau$ (including both control pulses and free evolution periods), the $l$-th pulse has a width $\tau_{p_{l}}$ and is centered at a time $\delta_{l}\tau$, where $0\leq\delta_{l}\leq 1$ and $\delta_{n+1}=1$ (see figure \ref{Fig:DDsequence}). Using this notation the sequence control vector is
\begin{eqnarray}\label{Eq:genDD}
\fl
\hspace{1cm}
\mathbf{R}_{z}(\omega)=\hat{\mathbf{z}}\left[1-e^{i\omega\tau}+\sum_{l=1}^{n}e^{i\omega\delta_{l}\tau}
\left(e^{i\omega\tau_{p_{l}}/2}\boldsymbol{\Lambda}^{(l)}
-e^{-i\omega\tau_{p_{l-1}}/2}\boldsymbol{\Lambda}^{(l-1)}\right)\right]\nonumber\\
\hspace{6cm}+\sum_{l=1}^{n}e^{i\omega\delta_{l}\tau-\tau_{p_{l}}/2}\mathbf{R}_{z}^{P_{l}}(\omega)
\boldsymbol{\Lambda}^{(l-1)}
\end{eqnarray}
the form of the pulse appearing explicitly only in the last term.

While DD sequences have traditionally been composed of identical $\pi$-pulses executed about a common axis, equation \eref{Eq:genDD} is quite general. The only restriction being that, in the absence of noise, the net effect is to execute the identity operation. Hence, it may be applied to the analysis of unusual DD schemes such as the recently proposed KDD (Knill dynamical decoupling), based on the Knill pulse \cite{Souza2011}. However, since our aim here is simply to provide an example application of our method, we will limit our attention to more conventional sequences. Specifically, we assume identical pulses, each of width $\tau_{p}$ and executing a net $\pi$-rotation of the qubit Bloch vector about the $x$ axis, i.e., $P_{l}=\sigma_{x}$, for $l\in\{1,2,...,n\}$. In this case, the pulse control vector $\bi{R}_{z}^{P}(\omega)$ is independent of $l$ and has only $z$ and $y$ components. The matrix $\boldsymbol{\Lambda}^{(l-1)}$ is diagonal with $\Lambda^{(l-1)}_{yy}$ and $\Lambda^{(l-1)}_{zz}$ alternating between $1$ and $-1$ with each successive pulse. The resulting sequence control vector has the two components
\numparts
\begin{eqnarray}\label{Eq:final1}
R_{zz}(\omega)=1-e^{i\omega\tau}+\left[2\cos{(\omega\tau_{p}/2)}-e^{-i\omega\tau_{p}/2}R_{zz}^{P}(\omega)\right]
\sum_{l=1}^{n}(-1)^{l}e^{i\omega\delta_{l}\tau}
\end{eqnarray}
and
\begin{eqnarray}\label{Eq:final2}
R_{zy}(\omega)=-e^{-i\omega\tau_{p}/2}R_{zy}^{P}(\omega)\sum_{l=1}^{n}(-1)^{l}e^{i\omega\delta_{l}\tau}
\end{eqnarray}
\endnumparts
It is worth noting here that, when the form of the pulse is ignored altogether (i.e., when both $R_{zy}^{P}(\omega)$ and $R_{zz}^{P}(\omega)$ go to zero), the control vector reduces to a single component
\begin{eqnarray}\label{Eq:nonoise}
R_{zz}(\omega)=1-e^{i\omega\tau}+2\cos{(\omega\tau_{p}/2)}
\sum_{l=1}^{n}(-1)^{l}e^{i\omega\delta_{l}\tau}.
\end{eqnarray}
a result derived previously by assuming finite-width pulses during which noise was ignored \cite{Biercuk2009a, Biercuk2009}. And, in the bang-bang limit of infinitely narrow pulses ($\tau_{p}\rightarrow0$) we have the standard result (for even $n$)
\begin{equation}\label{BBfilter}
R_{zz}(\omega)=1-e^{i\omega\tau}+2\sum_{l=1}^{n}(-1)^{l}e^{i\omega\delta_{l}\tau}.
\end{equation}
\noindent with $F_{z}(\omega)=|R_{zz}(\omega)|^{2}$.

Equations \eref{Eq:final1} and \eref{Eq:final1} capture modification of dephasing dynamics due to the duration and form of the pulses as well as depolarization effects occurring due to pulse-errors derived from a fluctuating detuning from resonance (i.e. the impact of dephasing noise). We will move forward using these expressions in order to treat experimentally relevant cases of DD incorporating various pulse forms including error-suppressing dynamically corrected gates.

\subsection{Pulse forms: Standard $\pi_{X}$-pulses and dynamically corrected gates}
For a DD sequence we are interested in physical control operations that implement a logical NOT gate.  The simplest approach is to employ ``primitive'' $\pi_{X}$-pulses, implemented by applying the ideal control Hamiltonian $H_{c}=\Omega\sigma_{x}/2$ over a time interval $[0,\tau_{\pi}]$, where $\Omega\equiv\pi/\tau_{\pi}$. This induces a simple rotation of the Bloch vector through an angle of $\pi$ radians about the $x$-axis.  In the presence of time-varying dephasing noise, the bare gate is modified by an error term that, to first order, results in a small additional rotation about an axis that has both $y$ and $z$-components. From equations \eref{Eq:seq2w} and \eref{Eq:pulsew} we find that the two corresponding components of control vector are
\numparts
\begin{equation}\label{Eq:primcv1}
R^{P}_{zz}\equiv R^{(Prim)}_{zz}(\omega)=\frac{\omega^{2}}{\omega^{2}-\Omega^{2}}\left(e^{i\omega\tau_{\pi}}+1\right)
\end{equation}
and
\begin{equation}\label{Eq:primcv2}
R^{P}_{zy}\equiv R^{(Prim)}_{zy}(\omega)=\frac{i\omega\Omega}{\omega^{2}-\Omega^{2}}\left(e^{i\omega\tau_{\pi}}+1\right)
\end{equation}
\endnumparts
It is interesting to note the correspondence between the prefactors here and those arising in a master-equation treatment of a driven quantum system in the presence of dissipation \cite{breuer2002theory}.

Moving beyond the standard $\pi_{X}$-pulse of finite duration, we consider a gate that has been designed to provide robustness in the presence of random gate errors - the dynamically corrected gate.  In previous work we demonstrated how such gates provide enhanced resistance to error in the presence of time-varying dephasing noise \cite{Green2012}.  Here we include expressions for a dynamically corrected logical NOT gate consisting of three successive $\pi_{X}$ pulses, the second of which is executed at a rotation rate $\Omega/2$, half that of the other two \cite{ViolashortDCG}. The control Hamiltonian for this gate is
\begin{eqnarray}
\label{cases}
H_{c}(t)=\cases{\Omega\sigma_{x}/2&$0\leq t<\tau_{\pi}$\\
\Omega\sigma_{x}/4&$\tau_{\pi}\leq t<3\tau_{\pi}$\\
\Omega\sigma_{x}/2&$3\tau_{\pi}\leq t\leq 4\tau_{\pi}$}
\end{eqnarray}
so that $\tau_{p}=4\tau_{\pi}$. In the case of dephasing noise, the control vector again has two components:
\numparts
\begin{eqnarray}\label{Eq:dcgv1}
R^{P}_{zz}\equiv R^{(DCG)}_{zz}(\omega)
=\omega^{2}\left[
\frac{p_{1}(\omega)}{\omega^{2}-\Omega^{2}}-\frac{p_{2}(\omega)}{\omega^{2}-(\Omega/2)^{2}}\right]
\end{eqnarray}
\begin{eqnarray}\label{Eq:dcgv2}
R^{P}_{zy}\equiv R^{(DCG)}_{zy}(\omega)
=i\omega\Omega\left[\frac{p_{1}(\omega)}{\omega^{2}-\Omega^{2}}-\frac{p_{2}(\omega)/2}{\omega^{2}-\left(\Omega/2\right)^{2}}
\right]
\end{eqnarray}
\endnumparts
where $p_{1}(\omega)\equiv e^{4i\omega\tau_{\pi}}+e^{3i\omega\tau_{\pi}}+e^{i\omega\tau_{\pi}}+1$ and $p_{2}(\omega)\equiv e^{3i\omega\tau_{\pi}}+
e^{i\omega\tau_{\pi}}$.

\subsection{Interpreting the filter function}
The inset to figure \ref{Fig:FF}a shows the filter functions, $F_{z}(\omega)\equiv |R^{P}_{zy}(\omega)|^2+|R^{P}_{zz}(\omega)|^2$,  for the primitive and dynamically corrected logical NOT operations described above, as a function of the angular frequency in units of $\tau_{\pi}^{-1}$. The rate of growth of the filter function with $\omega$, captures the low-frequency ``filtering'' performance of DD sequences.  Beyond a certain frequency $F_{z}(\omega)\to1$, above which noise is passed unimpeded.  This corresponds to the physical observation that fluctuations fast relative to the shortest interpulse period cannot be effectively suppressed by dynamical decoupling.

The filter function's behavior near zero frequency captures the leading-order error susceptibility.  We may express $F_{z}(\omega)\propto(\omega\tau_{\pi})^{2(\alpha+1)}$ as $\omega\tau_{\pi}\rightarrow 0$, then $\alpha$ is the order of error suppression for the sequence \cite{Hayes2011}.  Using these expressions, an examination of figure~\ref{Fig:FF} reveals that the relative performance of the primitive and dynamically corrected gates is captured in the slope of the filter function as plotted on a log-log scale~\cite{mbad2011}; higher slope corresponds to a higher order of error suppression (for noise near zero frequency).   In a filter-design framework we may refer to this slope as a filter rolloff in the stopband, using the relation that the rolloff is $6(\alpha+1)$ dB/octave. At low frequencies the primitive $\pi_{X}$-pulse filter function is approximately quadratic, i.e. $F_{z}(\omega)\propto(\omega\tau_{\pi})^{2}$, resulting in $\alpha=0$ - the same as for free evolution. This is not a surprising result, since there is no error suppression built into the primitive gate.  By contrast the DCG filter function scales as $F_{z}(\omega)\propto(\omega\tau_{\pi})^{4}$, giving $\alpha=1$, as expected for a construction designed to suppress error to first order.

\subsection{The full filter functions: DD + DCG}
We now combine the control vectors derived for the pulses themselves with those obtained for the generic DD sequences, considering three paradigmatic cases:  (1) bang-bang decoupling, (2) standard pulses with nonzero $\tau_{p}$, (3) DCGs.  This is accomplished by calculating $\bi{R}_{z}^{P}$ and inserting the relevant Cartesian components into~\eref{Eq:final1} and~\eref{Eq:final2} for the full DD sequence.  This study reveals how it is possible to evaluate complex perturbations to quantum control protocols in a straightforward manner using our method.

\begin{figure}[b]
      \centering
      \includegraphics[width=160mm]{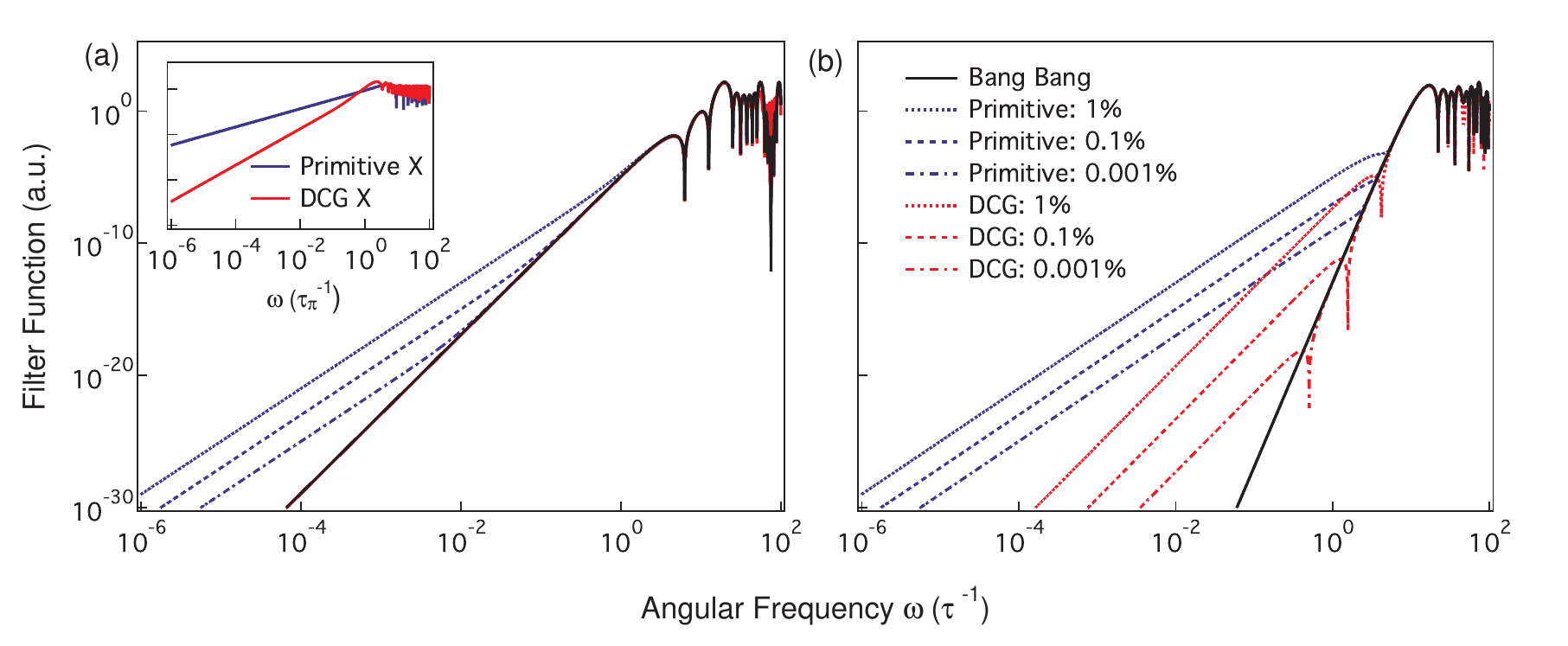}
      \caption{First order filter functions as a function of dimensionless frequency. The inset of a) compares filter functions for primitive (blue) and dynamically corrected (red) NOT gates while the main plot shows filter functions for a $6$ pulse CP sequence. The results for primitive NOT pulses are shown in blue, while those for dynamically corrected pulses are shown in red. Plot b) shows the same set of results for the $6$ pulse UDD sequence.  Black lines represent filter functions for sequences in the bang-bang limit}
      \label{Fig:FF}
\end{figure}

We compare the error-suppressing performance of CP (Carr and Purcell) and UDD (Uhrig dynamical decoupling) sequences as illustrative examples, accounting for each of the cases outlined above.  The CP sequence, devised in the context of NMR \cite{CP1954}, is a straightforward extension of the original Hahn spin echo \cite{Hahn1950}. In an $n$-pulse CP sequence, the $l$-th pulse has the fractional location $\delta_{l}=(l-1/2)/n$. The bang-bang limit CP filter function for $n=6$ is shown as a black line in figure~\ref{Fig:FF}a) for $n=6$. It exhibits an order of error suppression of $\alpha=2$, as do all CP sequences with $n\geq2$, giving a roll-off of 18 dB/octave.  This is a limiting case in which pulse effects are entirely ignored.

If we now take into account the width of the pulses by assuming that they take the form of primitive NOT gates of width $\tau_{p}$, as described above, then the components of the pulse control vector $\bi{R}_{z}^{P}(\omega)$ are given by equations \eref{Eq:primcv1} and \eref{Eq:primcv2}. The resulting filter functions, indicated by the blue lines in figure~\ref{Fig:FF}a, show a reduced order of error suppression $\alpha=2\to1$, with a corresponding reduction of the filter rolloff to $12$dB/octave.   As the pulse width grows as a fraction of the entire sequence, the relative importance of the pulses grows, as expected.  This is manifested as an expansion of the range over which the modification to the filter-function due to pulse effects dominates the bang-bang filter function.

Despite the reduction in the order of error suppression, there is still noise cancelation, regardless of the pulse width, since the sequence is in essence an Eulerian dynamical decoupling (EDD) sequence \cite{ViolaKnill2003}. Replacing the primitive pulses with dynamically corrected composite gates and the components of the pulse control vector with \eref{Eq:dcgv1} and \eref{Eq:dcgv2}, we almost completely restore the original $18$dB/octave roll-off for the CP sequences (figure~\ref{Fig:FF}a).  These filter functions lie approximately beneath the bang-bang filter function in this figure.  This observation confirms that the DCG provides one order of error cancellation during the applied pulses.  In essence, the deleterious effect of error accumulation during nonzero-duration control pulses can be mitigated by choosing a compensating pulse design.

Studying UDD is more instructive because the $n$-pulse UDD sequence, with relative pulse locations given by $\delta_{l}=\sin^2{[\pi l/(2n+2)]}$, ensures that (in the bang bang limit) the first $n$-derivatives of $R_{zz}(\omega)$ vanish at $\omega=0$~\cite{Uhrig2007, Uhrig2008}, giving an order of error suppression that increases linearly with $n$. The effects of pulse width and shape on a UDD sequence are therefore potentially more dramatic. For sufficiently long pulses we see, from figure~\ref{Fig:FF}b, that the general benefits derived from the use of a UDD sequence can largely be destroyed as the error-susceptibility of the pulses dominates the suppression provided by the pulse timing.  Again, we observe that the reduction in the order of error suppression arising from addition of nonzero-duration $\pi_{X}$-pulses can be partially offset by use of a DCG.  However, the fact that by design, the DCG used here only provides first-order error suppression, reduces our ability to recover the original order of error suppression provided by a UDD sequence with large $n$.

Nonetheless, these results indicate that judicious choice of DCG or other compensating pulse protocols within DD sequences provides a path to mitigate the effects of pulse errors in DD sequences.  This is especially important in long-storage settings where large pulse numbers may be employed in order to effectively suppress dephasing errors~\cite{Khodjasteh_LongTime}.  These observations represent an important validation of our method, and show that it provides a straightforward approach to account for a variety of pulse-duration, modulation, and shape effects.

\section{Limits of approximation}\label{Sec:Limits}

In deriving \eref{Eq:result2} in section \ref{Sec:Main}, we assumed both that the Magnus expansion converges and that truncating the additional series expansion of the fidelity \eref{Eq:fidexp} introduces no significant error. In this section we examine these assumptions more closely and discuss their limits of validity.

\subsection{Convergence of the Magnus expansion}
A sufficient condition for the convergence of the Magnus expansion is \cite{Moan1999}
\begin{equation}\label{meconv}
\int_{0}^{\tau}dt\|\tilde{H}_{0}(t)\|_{op}<\pi
\end{equation}
The operator norm $\|\tilde{H}_{0}(t)\|_{op}$ is the smallest number $K\geq 0$ such that $\|\tilde{H}_{0}(t)|\psi\rangle\|_{\mathcal{H}_{S}}\leq K\||\psi\rangle\|_{\mathcal{H}_{S}}$ for all $|\psi\rangle\in\mathcal{H_{S}}$, where $\||\psi\rangle\|_{\mathcal{H}_{S}}\equiv\sqrt{\langle\psi|\psi\rangle}$ is the usual vector norm defined on the system Hilbert space $\mathcal{H_{S}}$ \cite{Spindler}. It is a simple matter to derive the intuitively obvious result that
\begin{eqnarray}
\|\tilde{H}_{0}(t)\|_{op}=\|\boldsymbol{\beta}(t)\|=\left(\beta_{x}(t)^{2}+\beta_{y}^{2}(t)+\beta_{y}^{2}(t)\right)^{1/2}
\end{eqnarray}
i.e., the `size' of the toggling frame Hamiltonian is measured by the magnitude of the noise vector.

In an experiment, for each physical realization of the noise process, the $i$-th noise component $\beta_{i}(t)$ will have some maximum absolute value $\beta_{i}^{(m)}$, over the interval $[0,\tau]$. We can then write $\|\boldsymbol{\beta}(t)\|\leq
\|\boldsymbol{\beta}^{(m)}\|$, for all $t$, where $\boldsymbol{\beta}^{(m)}=(\beta^{(m)}_{x},\beta^{(m)}_{y},\beta^{(m)}_{z})$. It immediately follows that
\begin{equation}
\int_{0}^{\tau}dt\|\boldsymbol{\beta}(t)\|\leq
\|\boldsymbol{\beta}^{(m)}\|\tau
\end{equation}
and convergence is assured if $\|\boldsymbol{\beta}^{(m)}\|\tau<\pi$.
The problem is, of course, that the maxima $\beta^{(m)}_{i}$ are unknown quantities. Nonetheless, if the noise is stationary and has well-defined root mean square values $\delta\beta_{i}\equiv(\langle\beta_{i}(0)^2\rangle)^{1/2}$, for $i=x,y,z$, then for small total operation times $\tau$, letting $\beta_{i}^{(m)}=C_{m}\delta\beta_{i}$, for a sufficiently large value of $C_{m}>0$, will mean that contributions from elements of the ensemble not satisfying the convergence condition
\begin{equation}
\xi<\pi/C_{m}
\end{equation}
where $\xi\equiv \langle \boldsymbol{\beta}(0)\boldsymbol{\beta}(0)^{T}\rangle^{1/2}\tau$, can be ignored without significant error.

It can be shown (see Appendix) that this same parameter $\xi$ represents a bound on the strength of high-order terms in the expansion for the fidelity~\eref{Eq:fidexp} and serves as ``smallness parameter'' providing insight into when the first-order filter-function is sufficient to obtain a good approximation for the overall fidelity. Essentially, we require $\xi^2\ll1$ for the first order approximation to hold. In general, this is sufficient to guarantee convergence of the Magnus expansion for all but a statistically insignificant proportion of qubits in the ensemble.

\subsection{Numerical simulations and the first-order approximation}
\begin{figure}[b]
\centering
\includegraphics[width=135mm]{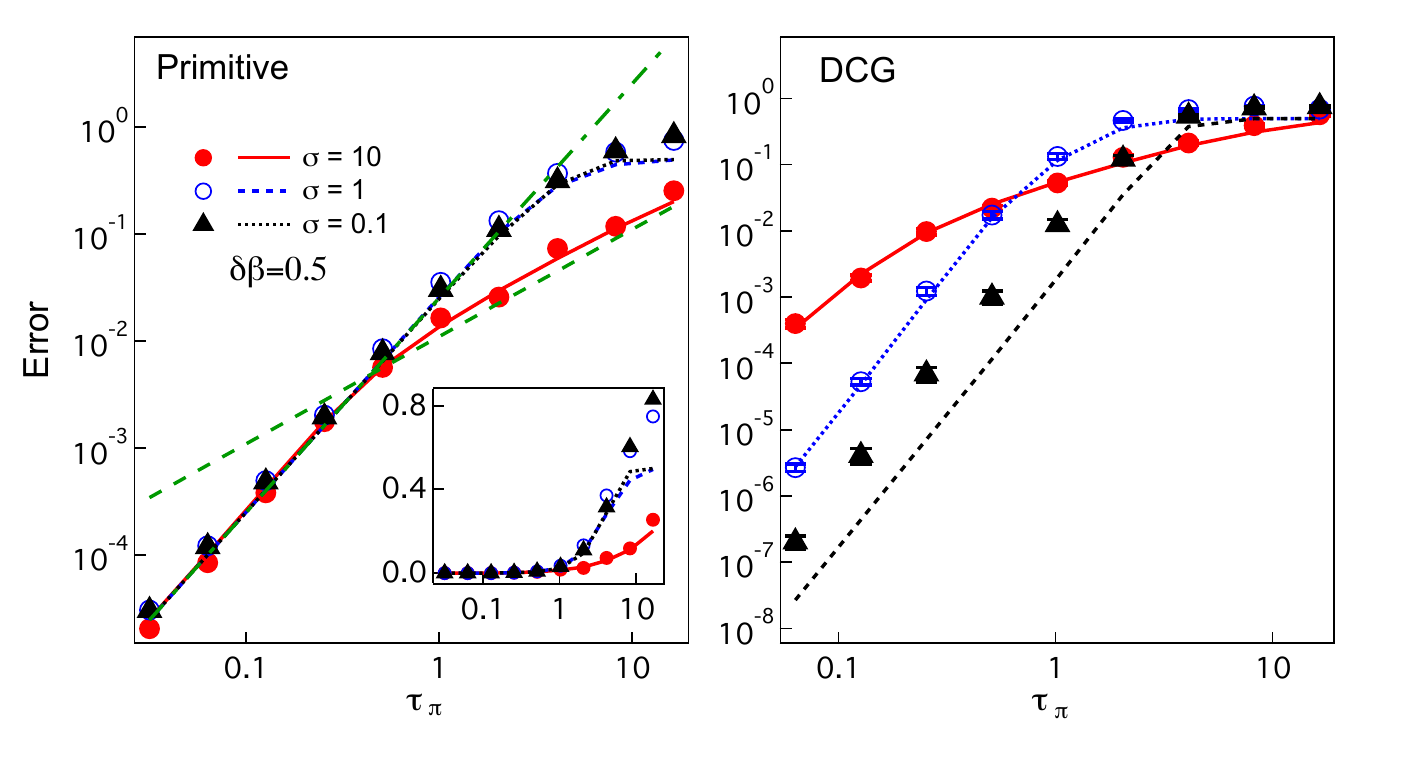}
\caption{Comparison of operational error derived from numerical simulations of Bloch vector evolution (markers) with analytical filter functions (lines) under application of a NOT ($\pi_{X}$-pulse), showing good agreement.  Left panel, Primitive gate, Right panel, Dynamically Corrected Gate.  Straight green dashed lines illustrate the scaling behaviour in the limiting regimes: Green dash-dotted line represents the approximate analytical behavior for $\tau_{\pi}/\tau_{c}\ll1$, green dashed line for $\tau_c/\tau_\pi\ll1$.  When conditions outlined in the text are not met, agreement between analytic and numeric results break down, as shown in the inset which plots the same data on a semilog scale. In this case, the origin of the breakdown is the growth of higher order terms in the expansion that are not accounted for in the lowest order expressions for the filter-function.}
\label{Fig:Num}
\end{figure}

Finally we present an instructive example to illustrate that our techniques are amenable to detailed understanding of the limits of applicability and scaling behaviours of the calculated operational fidelity. To this end we compare the performance of our analytical expressions with numerical simulations in which we
explicitly time-evolve the Schrodinger equation. As a test case, we simulate a primitive NOT-gate ($\pi_{X}$-pulse), in the presence of pure dephasing noise, $H(t)=\beta(t)\hat\sigma_z/2+\Omega_R\hat\sigma_x/2$, and study the behaviour of the fidelity \eref{Eq:fidexp} as a function of the control field strength.

For concreteness we pick a Gaussian noise power spectrum,
\begin{equation}
S_{z}(\omega)=\sqrt{2\pi}\frac{\delta\beta^2}{4\sigma}e^{-\omega^2/(2\sigma^2)}.
\end{equation}
Here $\delta\beta\equiv\delta\beta_{z}$, $\sigma$ is the bandwidth of the noise field, and the spectrum is properly normalized so that it is connected to the auto-correlation $C(t)$ of the noise through a Fourier transform:
\begin{equation}
C(t) = \frac{1}{2\pi}\int_{-\infty}^{\infty}S_{z}(\omega)e^{i\omega t}d\omega=\frac{\delta\beta^2}{4}e^{-t^2/(2\tau_c^2)},
\end{equation}
where $\tau_c=1/\sigma$.  For our simulations we vary the spectral bandwidth of the Gaussian spectrum.  The noise strength is parametrized by its root-mean-square deviation which we choose as $\delta\beta=0.5$ in each case. 

To simulate the noise, we generate random classical noise trajectories by summing noise Fourier
frequency components, each weighed by a random, normally distributed amplitude multiplied by the square root of the noise power at that frequency (as given by the power spectrum), and with a random phase. The gate fidelity is found by averaging over 100 realizations of the noise, and compared to the analytical
results including only first-order terms in the Magnus expansion. 

Analytic and numeric results are plotted in Fig.~\ref{Fig:Num}.  We observe excellent agreement between the first-order-filter-function analytical prediction and detailed numerical simulation.  We also find that there are appear to be two different regimes of behavior in relation to dependence on noise bandwidth (or identically the correlation time). The calculated gate fidelity appears to be independent of noise bandwidth in the short time regime $\tau_\pi\ll1$ and deviates from this behaviour only for the large bandwidth case $\sigma=10$.  We can understand this behaviour by analyzing in different temporal regimes, the behaviour of the first-order term in the fidelity expansion \eref{Eq:fidexp} which is simply 
\begin{eqnarray}
\langle a^2_1\rangle&=& \frac{1}{4}\int^{\tau_\pi}_0dt_2\int^{\tau_\pi}_0dt_1\delta\beta^2e^{-(t_1-t_2)^2/(2\tau_c^2)}\cos{(\Omega_Rt_1)}\cos{(\Omega_Rt_2)}\\
&+&  \frac{1}{4}\int^{\tau_\pi}_0dt_2\int^{\tau_\pi}_0dt_1\delta\beta^2e^{-(t_1-t_2)^2/(2\tau_c^2)}\sin{(\Omega_Rt_1)}\sin{(\Omega_Rt_2)}\\
&=& \frac{1}{4}\int^{\tau_\pi}_0dt_2\int^{\tau_\pi}_0dt_1\delta\beta^2e^{-(t_1-t_2)^2/(2\tau_c^2)}\cos{[\Omega_R(t_1-t_2)]}\label{Eq:a1primnot}
\end{eqnarray}
For short control times, i.e. $\tau_{\pi}/\tau_{c}\ll1$, this integral reduces to 
\begin{equation}\label{Eq:shorttime}
\langle a^2_1\rangle= \delta\beta^2\tau_\pi^2\left[\frac{1}{\pi^2}-\left(\frac{12-\pi^2}{4\pi^2}\right)\left(\frac{\tau_\pi}{\tau_{c}}\right)^2+\cdot\cdot\cdot\right]
\end{equation}
The lowest order term in Eq.~\ref{Eq:shorttime} is plotted as the green dash-dotted line in \ref{Fig:Num}. As demonstrated by both the analytical and the numerical results, one finds that the short-time behaviour is independent of bandwidth to lowest order.  

Bandwidth dependent effects should only appear when the control time becomes of the same order as the correlation time, reached for the case when $\sigma=10$.  To elucidate the behaviour in that regime we evaluate Eq.~\ref{Eq:a1primnot} approximately when $\tau_c/\tau_\pi\ll1$.  To lowest order this gives
\begin{equation}
\langle a^2_1\rangle\approx \sqrt{\pi}\frac{\delta\beta^2}{4}\tau_c\tau_\pi.
\end{equation}
This approximate solution is plotted as the green dashed line in the left panel of Fig.~\ref{Fig:Num}, illustrating that the full numerics and first-order analytics obey the expected scaling with control time, and that the error magnitude is set by the noise correlation time.  For a fixed root-mean-square noise strength, and long control times,  a primitive NOT gate is more robust under wide-bandwidth noise conditions.

Since $\xi^2=\delta\beta^2\tau_\pi^2\ll1$ over most of the regime we investigated, higher order terms in the Magnus expansion are expected to be negligible since they scale as $\xi^4$ to lowest order (see Appendix).  This expectation is confirmed by the numerics. The numerical results therefore support the scaling argument that $\xi^2\ll1$ determines the bounds of validity of the first-order filter-function approximation. The inset to Fig.~\ref{Fig:Num} shows that the analytical result becomes a poor approximation to the numerics when the condition $\xi^{2}\ll 1$ is violated in the vicinity of $\tau_{\pi}\approx10$.  Note that growth in higher-order terms does not necessarily coincide with divergence of the Magnus expansion, but rather a breakdown of the first-order approximation; higher-order filter functions may be incorporated to account for residual error.

Finally, we note that the same arguments we've applied to the primitive NOT gate may also be applied to more general control operations, including those designed to compensate for noise, such as DCGs.   First order error correcting gates like the DCG will reduce the effect of the first order term in the Taylor expansion of the noise correlation function relative to the higher-order bandwidth-dependent terms.  As a consequence, bandwidth effects are expected to be more pronounced for corrected gates, which we have confirmed via numerics. Despite this issue we still obtain good agreement between numerics and analytic filter functions to within factors of order unity, as demonstrated in previous work~\cite{Green2012}, and in the right panel of Fig.~\ref{Fig:Num}.

\section{Conclusion}
Understanding, predicting, and mitigating decoherence remain significant challenges for the development of quantum technologies.  In particular, in order to effectively develop novel quantum control protocols that suppress error or evaluate the performance of a complex quantum error correction procedure, we require methods permitting a system designer to estimate error probabilities using real inputs about the environment.  This capability is vital as the quantum science community moves from proof-of-principle concepts towards engineering of real quantum coherent technologies where rigid performance bounds are required.

 In this paper, we have presented a novel method for determining the fidelity of arbitrary quantum control operations in a universal time-dependent noise environment.  Building on past work employing spectral overlap functions, our method provides a straightforward, easily automated approach to producing analytic functions capturing the effect of the control in the Fourier domain -- generalized filter functions.  Using these we are able to evaluate control fidelities to arbitrary order, for arbitrary time-dependent control operations in terms of experimentally relevant characteristics of the noise.   While the error model we study is constrained to ignore, e.g., leakage errors from the qubit subspace or system-bath entanglement, the selected semiclassical model of universal, Non-Markovian, time-varying noise is far more realistic than earlier approximations based on uncorrelated errors.

We have exploited the capabilities provided by this technique to address the key challenge of accounting for nonidealities in control operations used to implement error-suppressing dynamical decoupling sequences.  By generating filter functions incorporating pulse errors and even the effect of complex intrapulse modulation schemes in a single compact formalism, we have provided a simple, physically intuitive means to evaluate sequence performance and inform the development of new dynamical error suppression protocols.

Our approach permits an experimentalist or system designer to analytically study the efficacy and error-susceptibility of customized control protocols involving complex temporal modulation schemes (e.g. control-field phase and amplitude).  For instance, it is possible to accurately account for error accumulation due to  \emph{control imperfections} arising from realistic constraints such as pulse overshoot, bandwidth limitations, or simple noise in the control field itself.  Further, for quantum information one may consider applying this approach at an algorithmic level, permitting ``echo'' type effects to be exploited in minimizing the error not just of a single operation, but of a complex chain in a computation.  With validation for the utility of our method from our study of dynamical decoupling and detailed numerics, we believe that this approach will prove invaluable for future experiments in which it is vital to accurately and efficiently estimate the effects of real experimental noise on the achievable fidelity of control operations.
\\
\\
\textit{Acknowledgements:}  The authors thank A.C Doherty, K. Khodjasteh, and L. Viola for useful discussions.  This work partially supported by the US Army Research Office under Contract Number  W911NF-11-1-0068, the Australian Research Council Centre of Excellence for Engineered Quantum Systems CE110001013, and the Office of the Director of National Intelligence (ODNI), Intelligence Advanced Research Projects Activity (IARPA), through the Army Research Office. All statements of fact, opinion or conclusions contained herein are those of the authors and should not be construed as representing the official views or policies of IARPA, the ODNI, or the U.S. Government.

\appendix

\section*{Appendix}

\setcounter{section}{1}

In this appendix we look more closely at the series expansion of the fidelity~\eref{Eq:fidexp}
\begin{equation}
\mathcal{F}_{av}=1-\langle a_{1}^{2}\rangle-2\langle \mathbf{a}_{1}\mathbf{a}^{T}_{2}\rangle-\left[\langle a_{2}^{2}\rangle+2\langle \mathbf{a}_{1}\mathbf{a}^{T}_{3}\rangle-\frac{\langle a_{1}^{4}\rangle}{3}\right]+...
\end{equation}
derived in section~\ref{Sec:Main} above. By calculating the maximum magnitude of the higher-order terms we can determine the conditions under which they can be neglected. We also show how, by converting to the frequency domain, the fidelity maybe expressed in terms of the noise power spectral density and a series of generalized filter functions.

\subsection*{Maximum magnitude of terms in the fidelity expansion}

In order to determine their relative contributions to the fidelity, we examine the terms in \eref{Eq:fidexp}, beginning with
\begin{eqnarray}\label{Eq:appfirstord}
\langle a_{1}^{2}\rangle&=\sum_{i,j=x,y,z}\int^{\tau}_{0}dt_{2}\int^{\tau}_{0}dt_{1}
\langle\beta_{i}(t_{1})\beta_{j}(t_{2})\rangle \bi{R}_{i}(t_{1})\bi{R}^{T}_{j}(t_{2})
\end{eqnarray}
We consider only the case in which the three noise components are independent Gaussian random processes, so that $\langle\beta_{i}(t_{1})\beta_{j}(t_{2})\rangle=\delta_{ij}\langle\beta_{i}(t_{1})\beta_{i}(t_{2})\rangle$ and
\begin{eqnarray}\label{Eq:appfirstord2}
\langle a_{1}^{2}\rangle&=\sum_{i=x,y,z}\int^{\tau}_{0}dt_{2}\int^{\tau}_{0}dt_{1}
\langle\beta_{i}(t_{1})\beta_{i}(t_{2})\rangle \bi{R}_{i}(t_{1})\bi{R}^{T}_{i}(t_{2})
\end{eqnarray}
Letting $\langle\beta_{i}(t_{1})\beta_{i}(t_{2})\rangle=\langle\beta_{i}^2(0)\rangle
\overline{\langle\beta_{i}(t_{1})\beta_{i}(t_{2})\rangle}$, where $\langle\beta_{i}^2(0)\rangle$ is the mean square value of the $i$-th component of the noise and
$|\overline{\langle\beta_{i}(t_{1})\beta_{i}(t_{2})\rangle}|\leq1$, we have
\begin{eqnarray}\label{Eq:appfirstord3}
|\langle a_{1}^{2}\rangle|&=\left|\sum_{i=x,y,z}\langle\beta_{i}^2(0)\rangle\int^{\tau}_{0}dt_{2}\int^{\tau}_{0}dt_{1}
\overline{\langle\beta_{i}(t_{1})\beta_{i}(t_{2})\rangle} \bi{R}_{i}(t_{1})\bi{R}^{T}_{i}(t_{2})\right|\nonumber\\
&\leq\sum_{i=x,y,z}\langle\beta_{i}^2(0)\rangle\left|\int^{\tau}_{0}dt_{2}\int^{\tau}_{0}dt_{1}
\overline{\langle\beta_{i}(t_{1})\beta_{i}(t_{2})\rangle} \bi{R}_{i}(t_{1})\bi{R}^{T}_{i}(t_{2})\right|\nonumber\\
&\leq\xi^2
\end{eqnarray}
where we've introduced the smallness parameter
\begin{eqnarray}
\xi^2\equiv\sum_{i=x,y,z}\langle\beta_{i}^2(0)\rangle\tau^2=\langle\boldsymbol{\beta}(0)\boldsymbol{\beta}^{T}(0)\rangle\tau^2
\end{eqnarray}

For Gaussian noise, correlation functions evaluated at an odd numbers of time points vanish, so
$
\langle \mathbf{a}_{1}\mathbf{a}^{T}_{2}\rangle=0.
$

Moving on to the first of the three terms that involve four-point noise correlation functions
\begin{eqnarray}
\fl\langle a_{2}^{2}\rangle=\sum_{i,j,i',j'}
\int_{0}^{\tau}dt_{1}\int_{0}^{t_{1}}dt_{2}
\int_{0}^{\tau}dt_{3}\int_{0}^{t_{3}}dt_{4}
\langle \beta_{i}(t_{1})\beta_{j}(t_{2})\beta_{i'}(t_{3})\beta_{j'}(t_{4})\rangle\nonumber\\
\hspace{7cm}\times\tilde{\bi{R}}_{ij}(t_{1},t_{2})\tilde{\bi{R}}^{T}_{i'j'}(t_{3},t_{4})
\end{eqnarray}
Here we can apply the Gaussian moment theorem to write
\begin{eqnarray}
\langle \beta_{i}(t_{1})\beta_{j}(t_{2})\beta_{i'}(t_{3})\beta_{j'}(t_{4})\rangle
=\langle \beta_{i}(t_{1})\beta_{j}(t_{2})\rangle\langle\beta_{i'}(t_{3})\beta_{j'}(t_{4})\rangle\nonumber\\
+\langle \beta_{i}(t_{1})\beta_{i'}(t_{3})\rangle\langle\beta_{j}(t_{2})\beta_{j'}(t_{4})\rangle
+\langle \beta_{i}(t_{1})\beta_{j'}(t_{4})\rangle\langle\beta_{i'}(t_{3})\beta_{j}(t_{2})\rangle
\end{eqnarray}
Using the independence of the noise components, this becomes
\begin{eqnarray}
\fl\langle \beta_{i}(t_{1})\beta_{j}(t_{2})\beta_{i'}(t_{3})\beta_{j'}(t_{4})\rangle
=\delta_{ij}\delta_{i'j'}\langle \beta_{i}(t_{1})\beta_{i}(t_{2})\rangle\langle\beta_{i'}(t_{3})\beta_{i'}(t_{4})\rangle\nonumber\\
\fl\hspace{2cm}+\delta_{ii'}\delta_{jj'}\langle \beta_{i}(t_{1})\beta_{i}(t_{3})\rangle\langle\beta_{j}(t_{2})\beta_{j}(t_{4})\rangle
+\delta_{ij'}\delta_{i'j}\langle \beta_{i}(t_{1})\beta_{i}(t_{4})\rangle\langle\beta_{i'}(t_{3})\beta_{i'}(t_{2})\rangle
\end{eqnarray}
so that
\begin{eqnarray}\label{Eq:appsec}
\langle a_{2}^{2}\rangle=\sum_{ij}\int_{0}^{\tau}dt_{1}\int_{0}^{t_{1}}dt_{2}\int_{0}^{\tau}dt_{3}\int_{0}^{t_{3}}dt_{4}\nonumber\\
\left\{\langle \beta_{i}(t_{1})\beta_{i}(t_{2})\rangle \langle\beta_{j}(t_{3})\beta_{j}(t_{4})\rangle
\tilde{\bi{R}}_{ii}(t_{1},t_{2})\tilde{\bi{R}}^{T}_{jj}(t_{3},t_{4})\right.\nonumber\\
\left.+\langle \beta_{i}(t_{1})\beta_{i}(t_{3})\rangle \langle\beta_{j}(t_{2})\beta_{j}(t_{4})\rangle
\tilde{\bi{R}}_{ij}(t_{1},t_{2})\tilde{\bi{R}}^{T}_{ij}(t_{3},t_{4})\right.\nonumber\\
\left.+\langle \beta_{i}(t_{1})\beta_{i}(t_{4})\rangle \langle\beta_{j}(t_{3})\beta_{j}(t_{2})\rangle
\tilde{\bi{R}}_{ij}(t_{1},t_{2})\tilde{\bi{R}}^{T}_{ji}(t_{3},t_{4})\right\}
\end{eqnarray}
and
\begin{eqnarray}\label{Eq:appa2norm}
|\langle a_{2}^{2}\rangle|\leq\sum_{ij}\langle\beta_{i}^2(0)\rangle\langle\beta_{j}^2(0)\rangle
\left|\int_{0}^{\tau}dt_{1}\int_{0}^{t_{1}}dt_{2}\int_{0}^{\tau}dt_{3}\int_{0}^{t_{3}}dt_{4}\nonumber\right.\\
\left.\left\{\overline{\langle \beta_{i}(t_{1})\beta_{i}(t_{2})\rangle} \overline{\langle\beta_{j}(t_{3})\beta_{j}(t_{4})\rangle}
\tilde{\bi{R}}_{ii}(t_{1},t_{2})\tilde{\bi{R}}^{T}_{jj}(t_{3},t_{4})\right.\right.\nonumber\\
\left.\left.+\overline{\langle \beta_{i}(t_{1})\beta_{i}(t_{3})\rangle} \overline{\langle\beta_{j}(t_{2})\beta_{j}(t_{4})\rangle}
\tilde{\bi{R}}_{ij}(t_{1},t_{2})\tilde{\bi{R}}^{T}_{ij}(t_{3},t_{4})\right.\right.\nonumber\\
\left.\left.+\overline{\langle \beta_{i}(t_{1})\beta_{i}(t_{4})\rangle} \overline{\langle\beta_{j}(t_{3})\beta_{j}(t_{2})\rangle}
\tilde{\bi{R}}_{ij}(t_{1},t_{2})\tilde{\bi{R}}^{T}_{ji}(t_{3},t_{4})\right\}\right|
\end{eqnarray}
Each of the three integrands in \eref{Eq:appa2norm} has a magnitude less than or equal to unity, so
\begin{equation}\label{Eq:appa1lim}
|\langle a_{2}^{2}\rangle|\leq3\sum_{ij}\langle\beta_{i}^2(0)\rangle\langle\beta_{j}^2(0)\rangle\tau^4/4
=3\xi^4/4
\end{equation}

Now
\begin{eqnarray}
\fl\hspace{1cm}\langle \mathbf{a}_{1}\mathbf{a}^{T}_{3}\rangle=\frac{2}{3}\sum_{i'ijk}\int_{0}^{\tau}dt_{1}\int_{0}^{\tau}dt_{2}
\int_{0}^{t_{2}}dt_{3}\int_{0}^{t_{3}}dt_{4}\langle \beta_{i'}(t_1)\beta_{i}(t_2)\beta_{j}(t_3)\beta_{k}(t_4)\rangle
\end{eqnarray}
which, using the Gaussian moment theorem and the independence of the noise components, we have
\begin{eqnarray}\label{Eq:appa1a3}
\langle\mathbf{a}_{1}\mathbf{a}^{T}_{3} \rangle=\frac{2}{3}\sum_{ij}\int_{0}^{\tau}dt_{1}\int_{0}^{\tau}dt_{2}\int_{0}^{t_{2}}dt_{3}\int_{0}^{t_{3}}dt_{4}\nonumber\\
\left\{\langle \beta_{i}(t_{1})\beta_{i}(t_{2})\rangle \langle\beta_{j}(t_{3})\beta_{j}(t_{4})\rangle
\bi{R}_{i}(t_{1})\tilde{\bi{R}}^{T}_{ijj}(t_{2},t_{3},t_{4})\right.\nonumber\\
\left.+\langle \beta_{i}(t_{1})\beta_{i}(t_{3})\rangle \langle\beta_{j}(t_{2})\beta_{j}(t_{4})\rangle
\bi{R}_{i}(t_{1})\tilde{\bi{R}}^{T}_{jij}(t_{2},t_{3},t_{4})\right.\nonumber\\
\left.+\langle \beta_{i}(t_{1})\beta_{i}(t_{4})\rangle \langle\beta_{j}(t_{3})\beta_{j}(t_{2})\rangle
\bi{R}_{i}(t_{1})\tilde{\bi{R}}^{T}_{jji}(t_{2},t_{3},t_{4})\right\}
\end{eqnarray}
Following the same procedure used to derive \eref{Eq:appa1lim}, we find that
\begin{equation}
|\langle\mathbf{a}_{1}\mathbf{a}^{T}_{3}\rangle|\leq\xi^{4}/3
\end{equation}

Finally,
\begin{eqnarray}
\langle a_{1}^{4}\rangle&=\sum_{i,j,i',j'}
\int_{0}^{\tau}dt_{1}\int_{0}^{\tau}dt_{2}
\int_{0}^{\tau}dt_{3}\int_{0}^{\tau}dt_{4}
\langle \beta_{i}(t_{1})\beta_{j}(t_{2})\beta_{i'}(t_{3})\beta_{j'}(t_{4})\rangle\nonumber\\
&\hspace{2cm}\left(\bi{R}_{i}(t_{1})\bi{R}^{T}_{j}(t_{2})\right)\left(\bi{R}_{i'}(t_{3})\bi{R}^{T}_{j'}(t_{4})\right)
\end{eqnarray}
which becomes
\begin{eqnarray}\label{Eq:appa4}
\langle a_{1}^{4}\rangle=\sum_{ij}\int_{0}^{\tau}dt_{1}\int_{0}^{\tau}dt_{2}\int_{0}^{\tau}dt_{3}\int_{0}^{\tau}dt_{4}\nonumber\\
\left\{\langle \beta_{i}(t_{1})\beta_{i}(t_{2})\rangle \langle\beta_{j}(t_{3})\beta_{j}(t_{4})\rangle
\left(\bi{R}_{i}(t_{1})\bi{R}^{T}_{i}(t_{2})\right)\left(\bi{R}_{j}(t_{3})\bi{R}^{T}_{j}(t_{4})\right)\right.\nonumber\\
\left.+\langle \beta_{i}(t_{1})\beta_{i}(t_{3})\rangle \langle\beta_{j}(t_{2})\beta_{j}(t_{4})\rangle
\left(\bi{R}_{i}(t_{1})\bi{R}^{T}_{j}(t_{2})\right)\left(\bi{R}_{i}(t_{3})\bi{R}^{T}_{j}(t_{4})\right)\right.\nonumber\\
\left.+\langle \beta_{i}(t_{1})\beta_{i}(t_{4})\rangle \langle\beta_{j}(t_{3})\beta_{j}(t_{2})\rangle
\left(\bi{R}_{i}(t_{1})\bi{R}^{T}_{j}(t_{2})\right)\left(\bi{R}_{j}(t_{3})\bi{R}^{T}_{i}(t_{4})\right)\right\}
\end{eqnarray}
We can then show that
\begin{equation}
|\langle a_{1}^{4}\rangle|\leq3\xi^{4}
\end{equation}
Obviously, for Gaussian noise, the overall trend here is that those terms containing $n$-point correlation functions make a maximum contribution of the order of $\xi^n$ to the fidelity. If we limit our attention to a regime in which $\xi^2\ll1$, then higher order terms will have little effect and the fidelity may be expressed in terms of the first
order error vector only.

\subsection*{Spectral representation}

The noise power spectral density $S_{i}(\omega)$ of the $i$-th noise component may be defined by
\begin{equation}\label{Eq:appPSD}
\langle\beta_{i}(t_{1})\beta_{i}(t_{2})\rangle=\frac{1}{2\pi}\int_{-\infty}^{\infty}d\omega
S_{i}(\omega)e^{i\omega(t_{2}-t_{1})}
\end{equation}
Noting that
\begin{equation}
\langle\beta_{i}^2(0)\rangle=\frac{1}{2\pi}\int_{-\infty}^{\infty}d\omega
S_{i}(\omega)=\frac{1}{\pi}\int_{0}^{\infty}d\omega
S_{i}(\omega)
\end{equation}
the smallness parameter $\xi$ may be expressed in terms of the noise spectrum. In particular,
if the $i$-th noise component has a cutoff frequency $\omega_{ci}$, then
\begin{equation}
\langle\beta_{i}^2(0)\rangle=\frac{1}{\pi}\int_{0}^{\omega_{ci}}d\omega
S_{i}(\omega)
\end{equation}
and
\begin{eqnarray}
\xi^2\equiv\frac{\tau^{2}}{\pi}\sum_{i=x,y,z}\int_{0}^{\omega_{ci}}d\omega S_{i}(\omega)
\end{eqnarray}

The dependence of the fidelity on the spectral properties of the noise and of the control can also be made explicit using \eref{Eq:appPSD}.
Substituting \eref{Eq:appPSD} into \eref{Eq:appfirstord2}, we have
\begin{eqnarray}\label{Eq:appfirstordps}
\langle a_{1}^{2}\rangle&=\frac{1}{2\pi}\sum_{i=x,y,z}\int_{-\infty}^{\infty}\frac{d\omega}{\omega^2}S_{i}(\omega)F^{1}_{i}(\omega)
\end{eqnarray}
where
\begin{eqnarray}\label{Eq:appfiltfunc}
F^{(1)}_{i}(\omega)\equiv\bi{R}_{i}(\omega)\bi{R}_{i}^{T*}(\omega)
\end{eqnarray}
is the first order generalized filter function for the $i$-th noise component, defined in terms of the $i$-th row of
the frequency domain control matrix
\begin{eqnarray}\label{Eq:appfiltfunc2}
\bi{R}_{i}(\omega)\equiv-i\omega\int_{0}^{\tau}dte^{i\omega t}\bi{R}_{i}(t)
\end{eqnarray}

Similarly, from \eref{Eq:appsec}
\begin{eqnarray}
\fl\hspace{1.5cm}\langle a_{2}^{2}\rangle=\frac{1}{(2\pi)^2}\sum_{ij}\int_{-\infty}^{\infty}d\omega S_{i}(\omega)
\int_{-\infty}^{\infty}d\omega' S_{j}(\omega')\int_{0}^{\tau}dt_{1}\int_{0}^{t_{1}}dt_{2}\int_{0}^{\tau}dt_{3}\int_{0}^{t_{3}}dt_{4}\nonumber\\
\fl\hspace{1.0cm}\left\{e^{i\omega(t_{2}-t_{1})}e^{i\omega'(t_{4}-t_{3})}\tilde{\bi{R}}_{ii}(t_{1},t_{2})\tilde{\bi{R}}^{T}_{jj}(t_{3},t_{4})
+e^{i\omega(t_{3}-t_{1})}e^{i\omega'(t_{4}-t_{2})}\tilde{\bi{R}}_{ij}(t_{1},t_{2})\tilde{\bi{R}}^{T}_{ij}(t_{3},t_{4})\right.\nonumber\\
\fl\hspace{2.5cm}\left.+e^{i\omega(t_{4}-t_{1})}e^{i\omega'(t_{3}-t_{2})}\tilde{\bi{R}}_{ij}(t_{1},t_{2})\tilde{\bi{R}}^{T}_{ji}(t_{3},t_{4})\right\}
\end{eqnarray}
while, from \eref{Eq:appa1a3}
\begin{eqnarray}
\fl\hspace{1.5cm}\langle\mathbf{a}_{1}\mathbf{a}^{T}_{3} \rangle=\frac{1}{(2\pi)^2}\sum_{ij}\int_{-\infty}^{\infty}d\omega S_{i}(\omega)
\int_{-\infty}^{\infty}d\omega' S_{j}(\omega')\int_{0}^{\tau}dt_{1}\int_{0}^{\tau}dt_{2}\int_{0}^{t_{2}}dt_{3}\int_{0}^{t_{3}}dt_{4}\nonumber\\
\fl\hspace{1.0cm}\left\{e^{i\omega(t_{2}-t_{1})}e^{i\omega'(t_{4}-t_{3})}\bi{R}_{i}(t_{1})\tilde{\bi{R}}^{T}_{ijj}(t_{2},t_{3},t_{4})
+e^{i\omega(t_{3}-t_{1})}e^{i\omega'(t_{4}-t_{2})}\bi{R}_{i}(t_{1})\tilde{\bi{R}}^{T}_{jij}(t_{2},t_{3},t_{4})\right.\nonumber\\
\fl\hspace{2.5cm}\left.+e^{i\omega(t_{4}-t_{1})}e^{i\omega'(t_{3}-t_{2})}\bi{R}_{i}(t_{1})\tilde{\bi{R}}^{T}_{jji}(t_{2},t_{3},t_{4})\right\}
\end{eqnarray}
and from \eref{Eq:appa4}
\begin{eqnarray}
\fl\hspace{1.5cm}\langle a_{1}^{4}\rangle=\frac{1}{(2\pi)^2}\sum_{ij}\int_{-\infty}^{\infty}d\omega S_{i}(\omega)
\int_{-\infty}^{\infty}d\omega' S_{j}(\omega')\int_{0}^{\tau}dt_{1}\int_{0}^{\tau}dt_{2}\int_{0}^{\tau}dt_{3}\int_{0}^{\tau}dt_{4}\nonumber\\
\fl\hspace{2.5cm}\left\{e^{i\omega(t_{2}-t_{1})}e^{i\omega'(t_{4}-t_{3})}\left(\bi{R}_{i}(t_{1})\bi{R}^{T}_{i}(t_{2})\right)\left(\bi{R}_{j}(t_{3})\bi{R}^{T}_{j}(t_{4})\right)\right.\\
\left.+e^{i\omega(t_{3}-t_{1})}e^{i\omega'(t_{4}-t_{2})}\left(\bi{R}_{i}(t_{1})\bi{R}^{T}_{j}(t_{2})\right)\left(\bi{R}_{i}(t_{3})\bi{R}^{T}_{j}(t_{4})\right)\right.\nonumber\\
\fl\hspace{2.5cm}\left.+e^{i\omega(t_{4}-t_{1})}e^{i\omega'(t_{3}-t_{2})}\left(\bi{R}_{i}(t_{1})\bi{R}^{T}_{j}(t_{2})\right)\left(\bi{R}_{j}(t_{3})\bi{R}^{T}_{i}(t_{4})\right)\right\}
\end{eqnarray}
Extrapolating these results, we find the following alternative form for the series expansion of the fidelity
\begin{eqnarray}\label{Eq:appfidspec}
\fl
\hspace{1cm}
\mathcal{F}_{av}=1-\left[\frac{1}{2\pi}\sum_{i}\int_{-\infty}^{\infty}\frac{d\omega}{\omega^2}S_{i}(\omega)F^{(1)}_{i}(\omega)\nonumber\right.\\
\hspace{1.6cm}\left.+\frac{1}{(2\pi)^{2}}\sum_{ij}\int_{-\infty}^{\infty}\frac{d\omega}{\omega^2}S_{i}(\omega)\int_{-\infty}^{\infty}\frac{d\omega'}{\omega'^2}
S_{j}(\omega')F^{(2)}_{ij}(\omega,\omega')+...\right]
\end{eqnarray}
where the generalized filter functions $F^{(p)}_{i_{1}i_{2}...}(\omega, \omega'...)$ are derived solely from the control matrix.

\section*{References}
\bibliography{njp5}

\end{document}